\documentclass{sig-alternate}
\usepackage{eurosym}
\usepackage{framed}
\usepackage{kpfonts}
\usepackage{xspace}
\usepackage{colortbl}
\usepackage[T1]{fontenc}
\usepackage[bottom]{footmisc}
\usepackage{mathtools}
\usepackage{listings}
\usepackage{cleveref}
\usepackage{enumitem}
\usepackage{accents}

\usepackage{booktabs}

\usepackage{graphicx}
\usepackage{caption}
\usepackage{subcaption}

\usepackage{color}
\usepackage[usenames,dvipsnames]{xcolor}
\usepackage[normalem]{ulem}

\usepackage{balance}
 \usepackage[]{algorithm2e}

\newtheorem{defi}{Definition}
\newtheorem{problem}{Problem}
\newtheorem{example}[defi]{Example}

\makeatletter
\def\@copyrightspace{\relax}
\makeatother

\def\pprw{8.5in}
\def\pprh{11in}

\usepackage[bookmarks,backref=true,linkcolor=black]{hyperref} 
\hypersetup{
  pdfauthor = {},
  pdftitle = {},
  pdfsubject = {},
  pdfkeywords = {},
  colorlinks=true,
  linkcolor= black,
  citecolor= black,
  pageanchor=true,
  urlcolor = black,
  plainpages = false,
  linktocpage
}

\definecolor{mred}{rgb}{.80,.12,.30}
\definecolor{grey}{rgb}{0.5,0.5,0.5}
\definecolor{Purple}{rgb}{.75,0,.85}
\definecolor{light-gray}{gray}{0.95}
\definecolor{mid-gray}{gray}{0.85}
\definecolor{darkred}{rgb}{0.7,0.25,0.25}
\definecolor{darkgreen}{rgb}{0.15,0.55,0.15}
\definecolor{darkblue}{rgb}{0.1,0.1,0.5}
\definecolor{blue}{rgb}{0.19,0.58,1}

\newtheorem{definition}{Definition}

\newcommand{\stitle}[1]{\smallskip\noindent\textbf{#1}}

\newcommand{\difftable}{\texttt{diffs}\xspace}
\newcommand{\difftablens}{\texttt{diffs}}
\newcommand{\diffspil}{\texttt{diffs\_pil}\xspace}
\newcommand{\diffspilns}{\texttt{diffs\_pil}}
\newcommand{\sys}{Precision Interfaces\xspace}
\newcommand{\lang}{\textsc{PILang}\xspace}

\DeclareMathOperator*{\argmax}{arg\,max}

\newif\ifnotes
\notestrue

\begin{document}

\title{Mining Precision Interfaces From Query Logs}
\subtitle{Technical Report}
\numberofauthors{3} 
\author{
\alignauthor
Haoci Zhang\\
       \affaddr{Columbia University}\\
       \email{hz2450@columbia.edu}
\alignauthor
Thibault Sellam\\
       \affaddr{Columbia University}\\
       \email{tsellam@cs.columbia.edu}
\alignauthor
Eugene Wu\\
       \affaddr{Columbia University}\\
       \email{ewu@cs.columbia.edu}
}

\maketitle

\begin{abstract}
  Interactive tools make data analysis both more efficient and more accessible to a broad population.  Simple interfaces such as Google Finance as well as complex visual exploration interfaces such as Tableau are effective because they are tailored to the desired user tasks. Yet, designing interactive interfaces requires technical expertise and domain knowledge. Experts are scarce and expensive, and therefore it is currently infeasible to provide tailored (or {\it precise}) interfaces for every user and every task.

We envision a data-driven approach to generate tailored interactive interfaces. We observe that interactive interfaces are designed to express sets of programs; thus, samples of programs---increasingly collected by data systems---may help us build interactive interfaces.  Based on this idea, {\it Precision Interfaces} is a language-agnostic system that examines an input query log, identifies how the queries {\it structurally change}, and generates interactive web interfaces to express these changes.  The focus of this paper is on applying this idea towards logs of structured queries.  Our experiments show that \sys can support multiple query languages (SQL and SPARQL), derive Tableau's salient interaction components from OLAP queries, analyze $<75k$ queries in $<12$ minutes, and generate interaction designs that improve upon existing interfaces and are comparable to human-crafted interfaces.

\end{abstract}

\section{Introduction}
\label{sec:intro}

Data analysis is a fundamental driver of modern decision making, and interactive interfaces are a powerful way for users to express their analyses. A well-designed interface provides interaction components for the users to easily accomplish their tasks and hides the technical complexity of the underlying system.  For instance, the Google Finance stock trend visualization incorporates a time-range filter so that users from a broad audience can explore prices over time. The same interface would not satisfy market analysts who want to aggregate sales information and perform roll-ups and drill-downs; for them, a rich interface such as Tableau would be preferable.  In a more extreme scenario, a single text box that lets users type a full program would be effective for engineers who want---and have the technical skills---to write all possible programs. This leads to the following observation:
\textbf{an interactive interface describes a set of programs, and its effectiveness depends on whether this set matches the operations that the users want to express.}

Designing interfaces poses two challenges traditionally addressed by experts:
{\it specifying} of a universe of programs that is relevant for a given task, and {\it developing} an interactive interface that can express it.
For example, consider Tableau~\cite{stolte2002polaris}, a popular visual data exploration tool. Tableau's designers carefully identified a common, and valuable, set of analytical data operations (OLAP) and mapped those to widgets in an interactive interface. Thus, they designed shelves to select measures or dimensions, and contextual menus to pick aggregates. Users did not need to learn SQL, and could easily manipulate queries by clicking, dragging and dropping interface components. Although this process was hugely successful, the cost---years of research and development---is not available for every task.

We believe that there is opportunity to drastically reduce the costs of interface construction through {\it data-driven} approaches.
Given a {\it trace of programs}---perhaps automatically logged by data processing systems---we may infer an interactive interface that can express them.
Although imperfect, this process can be largely automated, enabling us to scale interface construction and serve a long tail of users for whom it may not be feasible to manually build custom interfaces.
Our long term vision is to generate tailored interfaces for every user and every task based on their past analyses.


This paper is a first step towards this vision.  We propose {\it query logs} as the API for interactive interface generation because, increasingly, such logs are automatically collected by data processing systems by way of provenance capture subsystems~\cite{ground,scheidegger2008tackling}, as part of recovery and auditing mechanisms such as DBMS query logs~\cite{malviya2014rethinking}, or by user-facing applications such as Jupyter.
 Our discussions with several businesses have identified several compelling use cases for generating interfaces from those logs:

\stitle{Tailored dashboards:} An IOT startup (name anonymized) regularly performs tailored analyses for its customers. For simple cases, the engineers create custom front-ends with a dashboard builder. But the tool does not support complex statements (e.g., nested queries), and therefore the employees spend considerable time writing queries, including the CTO of the firm. For each case, they check out a text file that contains past queries, identify the statements that they need, customize them, copy-paste them and possibly update the document and check it in. A tool to build interfaces from queries would allow them to quickly set up expressive front-ends for each case and each customer.

\stitle{Auditing: } The employees of a large consulting firm perform financial audits by running previous programs (e.g., queries and macros) and performing what-if style analyses. In most cases, this task involves changing parameters, in order to test the robustness of financial indicators. But the auditors are not interested in programming.  \sys can be viewed as a way to summarize query logs into more accessible interactive visual interfaces.
\smallskip

To this end, we present \sys, an automatic tool to generate task-specific interactive interfaces from query logs.  \sys focuses on supporting different query languages\footnote{\small Our methods are based on abstract syntax trees, which we expect will generalize to different query languages.  However, our evaluation focuses on SQL and SPARQL queries.  We use ``program'' and ``query'' interchangeably in the text.} and different types of interface components (e.g., dropdowns, selections, panning, etc).  It takes a log as input (say, collected by IT) and generates a set of interactive web applications to express the queries in this log.  To do so, it parses the queries into canonicalized parse trees, compares pairs of trees to identify structural changes, and maps common types of changes to interface components.  Our focus is on identifying the salient interactions from query logs, and not necessarily the interface design per se.  Further, we are currently database agnostic and do not leverage information such as the schema, query plans, and the data.

Building \sys requires solving several key challenges.  First, we need to develop a unified mathematical model for queries and interfaces, which should be rich enough to express a wide range of real-life scenarios and incorporate user preferences but simple enough to remain tractable. Second, not all structural changes are meaningful and mapping all possible changes to interactions would lead to unusable interfaces---we need to devise a method to filter and identify the most important ones. Third, the space of all the interfaces that we can generate for a given set of query transformations grows exponentially with the size of this set. We need to define constraints and heuristics to quickly find good solutions and scale to logs that contain tens, or hundreds, of thousands of query. To tackle those challenges, we make the following contributions:
\begin{itemize}[leftmargin=*, topsep=0mm, itemsep=0mm]
\item We formalize the problem of mining structural changes in query logs and mapping them to interactive interfaces.  The problem definition is general to any language with a well-defined grammar, and a wide range of interaction components (including non-mouse interactions).
\item We introduce a DSL called \lang to specify interesting structural changes. This helps generate an interaction graph where each node is a query and each labeled edge is a structural change identified by a \lang statement.
\item We map the interface construction problem to edge cover over the interaction graph and use a contraction-based heuristic to quickly solve the problem.
\item We evaluate \sys{} on four query logs that span two query languages (SQL and SPARQL) and contains both synthetic and real-world statements. Our optimizations are 2OOM faster than the baseline, and lets us generate interfaces for logs of $>75k$ queries within 12 minutes.  The interfaces improve upon an existing interface that generated the queries, and are comparable with human crafted interfaces.

\end{itemize}

\begin{figure*}[t]
\centering
\includegraphics[width=\textwidth]{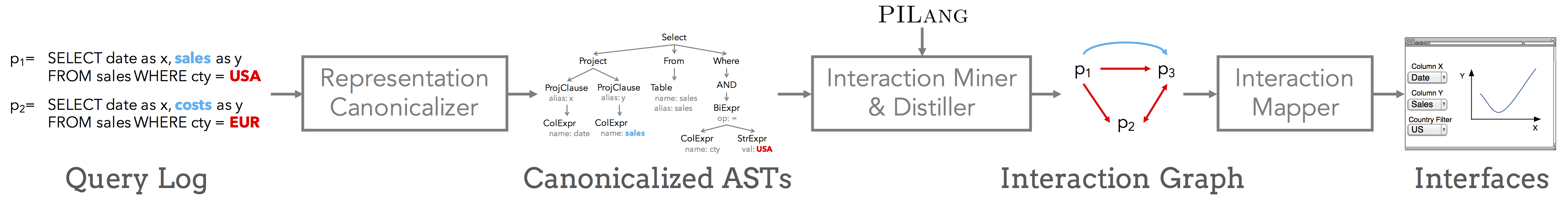}
\caption{\sys parses queries into canonicalized parse trees, performs tree alignment to generate an {\it interaction graph} that is filtered using the \lang domain specific language and  whose edges are mapped to interface widgets.}
\label{f:pipeline}
\end{figure*}

\section{Overview}
\label{sec:overview}

We give an overview of the precision interfaces setup and solution pipeline, as well as the technical challenges that we address in the subsequent sections.

\stitle{Query Logs as API} Interfaces are traditionally created by programmers or through a WYSIWYG application, so why mine interfaces from query logs?  The primary reason is that query logs encode the analyses that analysts {\it actually} perform, and therefore can be used to suggest candidate interfaces. As an API, logs are a flexible abstracton that can be generated from a variety of sources.  Modern program execution engines (e.g., DBMSs, Spark, Jupyter, RStudio) already track program logs for recovery and debugging purposes, while explicit provenance metadata systems are increasingly ubiquitous in industry~\cite{ground,mavlyutovdependency} and research~\cite{ives2008orchestra,muniswamy2006provenance,callahan2006vistrails}.  Any analysis that directly uses these systems, or uses them as a backend (e.g., Tableau) will naturally collect query logs.  Our hope is that if \sys is successful, then cleaning and finding query logs will be an interesting problem in its own right.


\subsection{Pipeline Overview}

We decompose the general problem of generating interfaces based on query logs in two sub-tasks: finding structural changes between queries and mapping those changes to interactions. The complexity and precision of the resulting interfaces depends on the types of structural changes that we can identify and the quality of the user interactions that we can map to arbitrary changes.

The problem is difficult because the scope of what a user interaction may express is theoretically unlimited---a button press could replace the current query with an arbitrary query string---and can easily lead to unusable interfaces.  We must bound the complexity of structural changes, and provide simple mechanisms to specify the types of changes that are meaningful.  Also, the system should easily adapt to new programming languages, as well as new types of interaction widgets (e.g., new modalities such as touch).

Based on these observations, we decompose the \sys process into three logical steps (Figure~\ref{f:pipeline}). The {\it Representation Canonicalizer} transforms the input sequence of query strings into a canonicalized parse tree structure (an AST) that makes query comparisons easier.

The {\it Interaction Miner and Distiller} logically identifies structural changes between the ASTs based on an ordered tree matching algorithm. These changes form an {\it interaction graph} where each node represents a query/AST and each directed edge is labeled by a corresponding structural transformation; it is a multi-graph because there may be multiple labeled edges between any two nodes.   To reduce the set of erroneous changes and preserve interesting changes, the developer can {\it distill} the graph by using a simple change specification language called \lang for specifying interesting structural changes.  We provide an interactive log analysis tool for suggesting potentially useful \lang statements, and use this tool in our experiment setup.

The {\it Interaction Mapper} maps sets of edges in the interaction graph to interaction components in interfaces.  Because this problem is NP-hard, we use a graph contraction heuristic to compute a best effort solution.  We then compile the resulting interfaces into an interactive web application.

\begin{figure}[ht]
\centering
\includegraphics[width=.4\columnwidth]{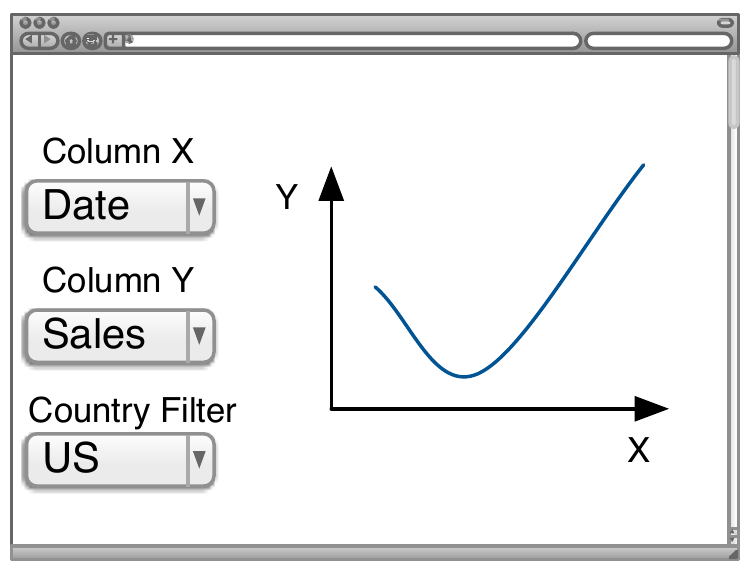}
\caption{An example interface.}
\label{f:dummy}
\end{figure}

%
%

\subsection{Challenges and Assumptions}
Real-life query logs may contain much variability, and it is not obvious how to map  arbitrary AST tree differences to widgets automatically. This leads to three major technical challenges.
The first is to develop a unified model of queries, interactions, interfaces and interface components (widgets) that is restricted enough for analysis (Section~\ref{s:model}).
But even then, the number of structural changes in a query log is quadratic in the log size and the majority of those changes are irrelevant. The second challenge is to develop mechanisms to identify the subset of changes that are meaningful to translate to interfaces (Section~\ref{s:pilang}).
Finally, the third challenge is to map these changes to components in one or more interfaces.  We show that this problem is NP-hard and present an efficient heuristics to generate the interfaces (Section~\ref{s:interface}).



\stitle{Inputs and Assumptions: } We assume that we have access to the grammar for each language, a parser to map program source code to a parse tree, an unparser to translate parser trees into source code~\cite{unparse}, and annotations of AST node types to understand which nodes are literals, or collections.

A core assumption is that most syntactic changes in the query log are {\it incremental}, such that the changes can be mapped to interactive interface components. To test this property, we evaluated \sys on query logs generated by students analyzing a dataset using SQL. We found that a majority of the queries changes conform the the assumptions, as we will show in Section~\ref{sec:case_stud}. In addition, we do not assume deep semantic understanding about the queries beyond near-universal features such as primitive data types---the whole analysis is performed syntactically.

Another assumption on which \sys{} is built is that there exists no logical dependency between the entries in the log -- for instance, no query uses a view or a temporary table defined in another statement. One way to remove this limitation would be to detect clusters of queries using, e.g., pattern mining or source code analysis. We leave this line of study for future work.

Finally, we assume two functions $exec()$ and $render()$ for a given language that executes a query AST and renders the output, respectively.  $exec()$ is called on an interface's current query state and for SQL query logs, $render()$ either generates a simple visualization~\cite{mackinlay2007show,mackinlay1986automating} or renders a table.



\section{Modeling Interactions}\label{s:model}

\begin{table}[b]
	\small
	\centering
	\begin{tabular}{rl}
			\toprule
			\textbf{Symbol}  & \textbf{Description}   \\
			\midrule
			$I$, $I_{closure}$  & Interface, its closure (expressible queries) \\
			$p \in P_{log}$  & Query $p$ in  query log $P_{log}$\\
			$\pi$, $\tau$  & Path in an AST, a subtree \\
			$t_\pi$ & Interaction that replaces subtree rooted at $\pi$ \\
			$\theta$, $w$  & Widget type and widget \\
			$C_\theta(), C_w()$  & Cost functions for widget type and widget \\
			$\Omega_\theta$, $\Omega_w$  & Domain for widget type and widget \\
			$f_w$  & Template function for widget $w$\\
			$\alpha_i$  & Cost function weight \\
			$s$ & \lang statement \\
			\bottomrule
			\\
	\end{tabular}
	\label{t:symbols}
	\caption{Summary of notations.}
\end{table}

This section describes the formal model of query logs and interactions on which we built \sys{}. The key idea is to model interactions as tree transformations, which serve to bridge the structural differences that we mine from the query logs with the interface components that users directly manipulate.

\subsection{Modeling Queries}\label{s:model-progs}

We assume that the input query log $P_{log}$ can be modeled as a table \texttt{queries} that contains the query id $pid$, parsed query $p$, along with any relevant metadata about the queries (e.g., the user that executed the program, the timestamp, the analysis session, etc):
{\small\begin{center}
  \texttt{queries(pid, p, tstamp, user, ...);}
  \end{center}
}
In order to support a variety of programming languages, we rely on the language's grammar and parse structure rather than query semantics.

\begin{figure}[h]
\centering
\includegraphics[width=\columnwidth]{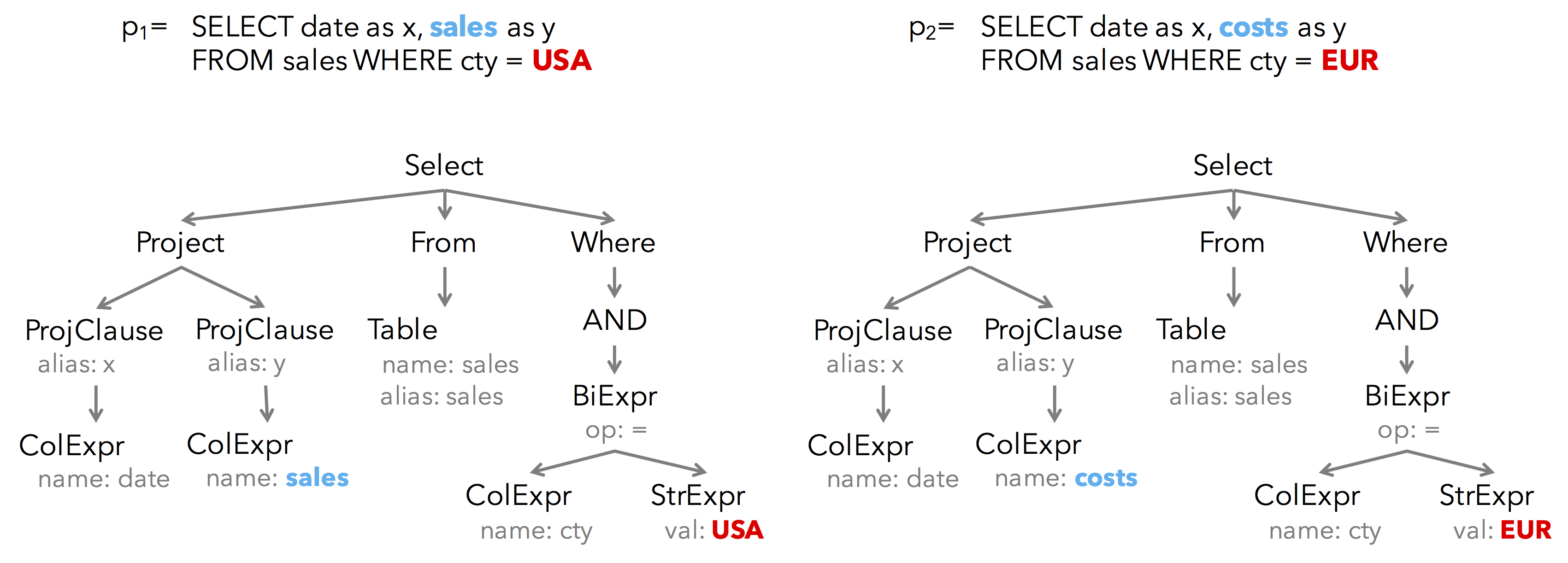}
\caption{\small Example ASTs for two SQL queries that differ in the second project clause (blue) and the constant in the equality predicate (red). }
\label{f:asts}
\end{figure}
We model a query $p$ as a canonicalized abstract syntax tree (AST) generated by a parser.  Figure~\ref{f:asts} shows two (simplified) examples of ASTs for two SQL queries.
We assume that each AST node consists of its node type, a set of attributes, and an ordered list of child nodes.  For instance the binary expression \texttt{cty = USA} is represented by the \texttt{BiExpr} node type, its attribute \texttt{op:=}, and two children for the left and right sub-expressions.  Its second child is a string literal \texttt{StrExpr} with value \texttt{USA}.

In addition, we assume the existence of a table that indicate how terminal nodes map to primitive data types (e.g., string literals map to \texttt{StrExpr}, integers map to \texttt{IntExpr}), as well as the node types that represent lists of sub-expressions (e.g., \texttt{Project} consists of a list of \texttt{ProjectClause} nodes).  Primitive data types are straightforward to identify as literal expressions in the language grammar.  Similarly, list nodes can either be identified by a language expert, or determined automatically by searching the language's grammar file for common idioms.  For instance, the SQLite grammar defines the list of output expressions \texttt{sel\_core} as a project clause (represented by the \texttt{sel\_result} non-terminal) followed by zero or more additional project clauses:
{\small
  \begin{verbatim}
  sel_core = (sel_result (whitespace comma sel_result)*)\end{verbatim}
}
Explicitly modeling list nodes lets \sys map multi-selection widgets, which can specify sets of values, to collection-based tree modifications that insert, reorder, or delete multiple AST subtrees.  For instance, a checkbox list of table attributes could be used to specify the list of attributes to return in the project clause.

\begin{figure}[b]
\centering
\includegraphics[width=.9\columnwidth]{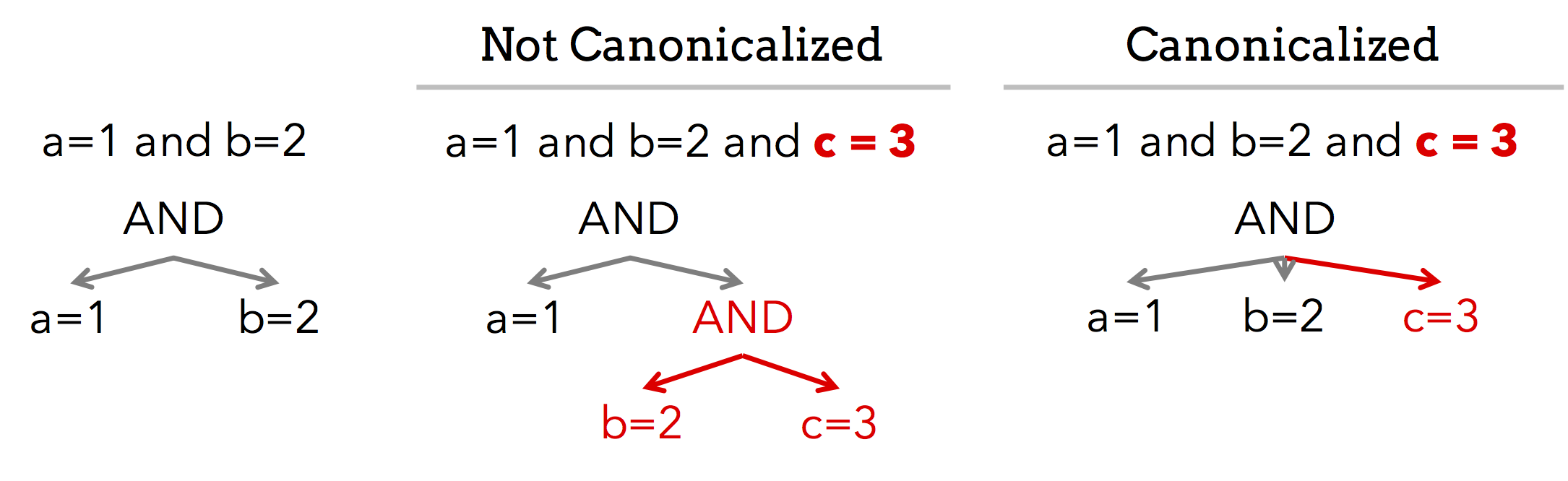}
\caption{Examples of non/canonicalized versions of ASTs for a SQL \texttt{WHERE} clause with two and three predicates. Canonicalization flattens the tree.}
\label{f:canonicalize}
\end{figure}

Finally, we assume that logical expressions are {\it canonicalized} into conjunctive normal form.  This allows us to model logical expressions as a list of lists (e.g., \texttt{AND}s of \texttt{OR}s) rather than a complex binary tree structure of \texttt{AND} and \texttt{OR} operators.  This reduces tree mis-alignment issues that can be caused when adding a logical expression to e.g., the \texttt{WHERE} clause of a query restructures the expression subtree (Figure~\ref{f:canonicalize}).

Although this process must be done for each language (and dialect), it only needs to be setup once (e.g., by an admin).  In this paper, the authors have provided drivers to model SQL and SPARQL queries.

\subsection{Interactions as Query Differences}
\label{sec:diff}
Given a list of ASTs, we would like to identify all structural differences
between pairs of trees.  To do so, we use a fast ordered tree matching
algorithm~\cite{bille2005survey,thomas2001introduction} that preserves ancestor
and left-to-right sibling relationships when matching nodes between the two
trees. The algorithm first computes the preorder traversal of both trees. It
goes on to the next node if the current pair of nodes matches. When the algorithm
finds a pair of nodes that cannot be mapped, it uses backtracking to return to
the last pair of nodes that has already been mapped and tries to map them to
some other candidate.  The algorithm has $O(\Pi_{i\in\{1,2\}} (T_i \times
min(L_i,D_i)))$ complexity where $T_i,L_i,D_i$ are respectively the size,
number of leaves, and the tree depth of the $i^{th}$ tree.

We model all pairwise AST differences in a set of queries $P$ as a logical table of differences \difftablens$_P$ that contains the sub-tree differences as output by a tree-alignment algorithm.  Specifically, \difftablens$_P$ contains foreign key references $pid_1$, $pid_2$ to the queries $p_1$ and $p_2$, the unique path $\pi$ to the sub-tree differences, and the sub-trees $\tau_1$ and $\tau_2$ that differ.
{\small
  \begin{center}
    \texttt{diffs($did$, $pid_1$, $pid_2$, $\pi$, $\tau_1$, $\tau_2$)}
  \end{center}
}
Additions and deletions in the ASTs can be represented by setting $\tau_1$ or $\tau_2$ to \texttt{null}, respectively. Let \difftable be shorthand for \difftablens$_{P_{log}}$.
\begin{example}\label{e:asts}
The pair of ASTs in Figure~\ref{f:asts} differ in the project clause and the equality predicate.  These differences would be modeled as two records in \difftable:
\begin{center}\small
\hspace*{-.1in}
        \begin{tabular}{ccclll}
            \toprule
            $id$ & \textbf{$pid_1$}  & \textbf{$pid_2$}    & \textbf{$\pi$}& \textbf{$\tau_1$}& \textbf{$\tau_2$}\\
            \midrule
            0 & 0 & 1 & 0/1/0   & \texttt{ColExpr(sales)} & \texttt{ColExpr(costs)} \\
            1 & 0 & 1 & 2/0/0/1 & \texttt{StrExpr(USA)} & \texttt{StrExpr(EUR)} \\
            \bottomrule
            \\
        \end{tabular}

\end{center}
The paths specify the index of each child along the path.  For instance, the first row's path follows the root node to \texttt{PROJECT} (0/), to the second \texttt{ProjClause} (0/1/), to its only child (0/1/0).  The transformation is to replace the \texttt{sales} column expression with \texttt{costs}.  The second row replaces the string expression \texttt{USA} with the string \texttt{EUR}.
\end{example}
Note that \difftable is a logical representation used to mine for interactions.   Our experiments show it is too costly to fully materialize \difftable for all but the smallest query logs.  In addition, not all structural differences are meaningful for generating interfaces.  Section~\ref{s:pilang} introduces a simple filtering language to identify meaningful subsets of \difftable.

\stitle{Interactions: } Interactions are the abstraction that connects records in the \difftable, which represent the locations and examples of structural changes in queries, with interface components, which translate user interactions into query transformations.  To this end, we model an interaction $t$ as a tree transformation function:
\begin{definition}
An interaction $t$ maps an AST $p$ into another AST $p'$ by substituting the subtree rooted at $\pi$ by a new subtree $\tau$:
  $t_{\pi}(p, \tau) = p'$
\end{definition}

\begin{example}
Continuing the example in Figure~\ref{f:asts}, the following interactions changes $p_1$'s project clause and then the equality predicate and outputs $p_2$:
$$p_2 = t_{2/0/0/1}(t_{0/1/0}(p_1, \texttt{ColExpr(costs)}), \texttt{StrExpr(EUR)}) $$
\end{example}

\stitle{Interaction Graph: } The table \difftable can be modeled as an {\it interaction graph}, where each query is a node, and a directed edge $p_i \xrightarrow{t_\pi, \tau} p_j$ is labeled with an interaction such that $p_j = t_\pi(p_i, \tau)$.  There can be multiple labeled edges between two nodes.  These two representations are interchangable, however the graph formulation is useful for generating interfaces in Section~\ref{s:interface}.

\subsection{Interfaces}

A given interactive interface $I = (p^I_0, W^I)$ represents the AST $p^I_0$ of an initial query  along with a set of interaction components $W^I$ such as buttons, sliders, selection, dragging, panning, and other manipulations that can be interactively expressed. In a slight abuse of terminology, we term these components {\it widgets}\footnote{\small Although we use the term widgets for simplicity, it also represents user manipulations such as panning that do not have a visual representation.}.  Each widget $w \in W$ incrementally transforms the current query $p^I_i$ into the next query $p^I_{i+1} = w(p^I_i)$, whose output is rendered in the interface. In effect, $I$ represents the set of queries expressible by applying all possible sequences of its widgets to its initial query $p^I_0$, which we term the closure $I_{closure}$ of the interface.  $exec()$ and $render()$ runs and renders the query AST.

We model a widget $w^\theta$ as an instance of a widget type $\theta$.  Consider the dropdowns in Figure~\ref{f:dummy}. A dropdown is a {\it type} of widget that renders a list of possible options that the user can select from, and its state stores the currently selected option; it is suitable for choosing from a small set of string options, and is more challenging to use when there are more than a dozen options.

 More generally, a widget type consists of a domain that restricts the allowable values $\Omega_\theta$, along with a generic cost function $C_\theta(\Omega) \in \mathbb{R}$ that quantifies how ``good'' the widget type is for a given domain $\Omega \subseteq \Omega_\theta$.  For instance, the general domain of a dropdown is the set of all possible strings, while for a range slider it is a pair of numbers $\{(v_{min}, v_{max}) \in \mathbb{R}^2 | v_{min} < v_{max}\}$.  Although these general domains are broadly defined, they are important for identifying candidate widgets that can express a given structural change.  We describe the cost function in the next subsection.

A widget $w^\theta = (t_{\pi}, \Omega_w, f_w)$ is an {\it instance} of a widget type $\theta$ that is instantiated with a specific domain $\Omega_w \subseteq \Omega_\theta$, as well as specifications of how the state of the widget should be used to modify a program.  The latter is specified by an interaction $t_{\pi}$ along with a {\it template function} $f_w(o) = \tau$ that maps an element $o \in \Omega_w$ to a subtree $\tau$ that can be passed as an argument to the interaction. Let $o_{w} \in \Omega_w$ be the current state of the widget; then applying the widget to the current query is equivalent to:
$$w(p) = t_{\pi}(p, f_w(o_w))$$

\begin{example}
Consider the interface in Figure~\ref{f:dummy}: it contains three dropdown widgets and its current query is the following SQL query whose output is rendered as a line chart: {\small\begin{center}
\texttt{SELECT date as x, sales as y FROM sales WHERE cty = 'US'}
\end{center}} The top \texttt{Column X} widget $w_{top}$ uses the selected value to modify the column expression \texttt{date} in the first projection clause; its domain is the set of attribute names in the table $\Omega_{w_{top}} = \{date, sales, \cdots\}$, its function $f(o) = ColExpr(o)$ returns a column expression populated with the specified attribute name, and its interaction replaces the subtree rooted at the first project clause with the output of $f(o_{w_{top}})$. Similarly, the middle widget sets the column expression of the second project clause, while the bottom widget modifies the string literal in the equality expression of the \texttt{WHERE} clause.
\end{example}
Our definition of widgets simply specifies a domain and widget state, and is not bound any specific form element.  This allows \sys to be easily extended to new interaction components or even different modalities such as voice or touch gestures.


\stitle{Invalid Queries: }  Since \sys operates at the syntactic level, certain combinations of AST transformations might lead to non-executable queries.  Although this is unlikely for common transformations such as adding expression clauses or tuning parameters, it is still possible.  One solution is to speculatively parse and execute queries in the interface's closure, and visually disallow interactions that lead to these ASTs.  If the space of queries is small, this can be a way to both verify and pre-compute results for performance purposes.    

\stitle{Ranking Interfaces:}
It is clear that there are many possible interfaces that could be used to express the same query log.  For instance, given a query log $P_{log}$, an interface may generate $|P_{log}|$ buttons, where each button widget $w_i$ represents query $p_i \in P_{log}$ and shows its result when pressed.  However, if all of the queries were identical except for a numerical constant that represents a threshold, then a single numerical slider would succinctly express the same set of queries.   Thus it is desirable to define a scoring function in order to rank and select the ``best'' interfaces.

The literature on assessing interactive interfaces is continuously evolving and has found a variety of characteristics that affect interface usability.   The GOMS family of interface analysis techniques  assign each user operation a cost and measure interface efficiency based on the cost to complete higher level goals~\cite{john1996goms,card1983psychology}.  Similarly, the amount of visual clutter~\cite{rosenholtz2007measuring} or even number of pixels needed to render the interface~\cite{peytchev2006web,jones1999improving} can affect readability. To flexibly support this range of interface measurements, we allow developers to specify multiple cost functions for the widget types.  As introduced in the previous subsection, the cost function for a widget type $\theta$ is defined as:
 $$C_\theta(\Omega) = \sum^k_{i=1} \alpha_{i} \times C^i_\theta(\Omega)$$
\noindent where $C^i_\theta(\Omega) \in [0, 1]$ is the $i^{th}$ cost function defined for the widget type by the developer.  We assume that each widget type has $k$ cost functions, whose outputs are between $[0, 1]$---a button may return $1$ if its domain contains more than one element, and $0$ otherwise: $max(0, min(1, |\Omega|-1))$.  Similarly, a checkbox list may increase linearly with the size of the domain: $min(1, \frac{|\Omega|}{12})$.  The $\alpha_{i}$ terms are user-controllable knobs to specify which cost functions matter more to the user.  For instance, in a setting with small screen resolution, the user may prioritize simpler interfaces that are easier to use than more complex and efficient interfaces that would necessitate scrolling or repeatedly zooming in~\cite{peytchev2006web}.

For an interface $I$, we estimate the interface complexity as weighed sum of its widgets:
$$C_I = \sum_{w^\theta \in W}  C_\theta(\Omega_w)$$

\stitle{Multiple Interfaces: }
In many cases, having \emph{one} interface that expresses all the queries in the log is not the optimal solution. Suppose for instance that our log contains only a pair of queries $\{p_0, p_1\}$ and that those queries are very different from each other. One approach is to create an interface $I=(p_0, \{w\})$ where $w$ is a widget that expresses the complex transformation between $p_0$ and $p_1$. Another approach is to create two interfaces $I_0=(p_0, \{\})$ and $I_0=(p_1, \{\})$, such that each interface expresses exactly one program.
To model this flexibility, we support sets of interfaces $\mathbb{I}$. We estimate that the complexity $C_\mathbb{I}$ as the sum of its interfaces:$C_\mathbb{I} = \sum_{I \in \mathbb{I}} (c_0 + C_I)$, where $c_0 \in [0, 1]$ is a constant cost for each new interface.   Similarly, we define the \emph{closure} $\mathbb{I}_{closure}$ as the union of its interface's closures: $\mathbb{I}_{closure} = \bigcup_{I \in \mathbb{I}}  I_{closure}$.

\subsection{Interface Generation Problem}\label{s:problem}
We can now define the main problem statement:
\begin{problem}[Interface Generation]
  Given a query log $P_{log}$, a threshold $\gamma$ for the percentage of the query log to cover, and the $\alpha_i$ weights for the cost functions, generate the optimal set of interfaces $\mathbb{I}^*$ such that:
  \begin{itemize}[topsep=-1mm, itemsep=0mm]
  \item $|\mathbb{I}^*_{closure} \cap P_{log}| \ge \gamma \times |P_{log}|$
  \item $C_{\mathbb{I}^*}$ is minimal
  \end{itemize}
\end{problem}

Our aim is to find a set the minimal interface which closure includes a given proportion $\gamma$ of the queries in the log.

\stitle{Solution Overview:} Our solution decomposes this problem in two steps.  The first is to efficiently mine the query log to identify meaningful structural changes that can be mapped to interactions; directly using the table \difftable can lead to overly complex and incoherent interfaces because it can contain both irrelevant differences as well as differences that syntactically appear similar but are semantically different.  To address this issue,  Section~\ref{s:pilang} introduces a domain specific language called \lang to filter \difftable.  An added benefit is that the filtering operations defined by \lang statements can be pushed into the query parsing and tree alignment steps of the system to improve the end-to-end runtime.

The second step is to map these changes to the appropriate widget types, and instantiate the widgets by generating each widget $w$'s domain $\Omega$, interaction $t_\pi$ and template function $f_w$.  To do so, Section~\ref{s:interface} describes how interface generation is modeled as a subset cover problem, and how to instantiate each widget from the changes output from the \lang statements.

\section{Interaction Mining}\label{s:pilang}

The Interface Generation problem states that the output interface should be capable of expressing the queries in the log.  However naively mining the log for all query differences leads to irrelevant differences (say, between two unrelated queries), as well as semantically similar but syntactically different changes (e.g., table aliases), that result in overly complex or semantically meaningless interactive interfaces.  In other words, not all possible interactions that can be mined from the query log are meaningful.    In this section, we present \lang, a domain specific language for specifying types of meaningful changes, as well as a tool to help developers write \lang statements.

\subsection{Why \lang?}

Why not use \difftable to analyze all differences in the log?
There are three types of issues that arise:

\stitle{Irrelevant Changes: } Changes to queries such as function renaming, changing the alias of a project clause, reordering tables in the \texttt{FROM} clause do not have any impact on the semantics of the query.

\stitle{Misleading Differences: } Consider the following two example queries:
{\small\begin{verbatim}
      SELECT a FROM T WHERE 1 = 1
      SELECT a FROM T GROUP BY a\end{verbatim}}
\noindent the tree alignment algorithm would identify that replacing the first query's \texttt{WHERE} clause with the \texttt{GROUPBY} clause produces the second query.  Even if this pair of queries are found in the query log, it is unlikely to be a meaningful interaction to map to a widget.  For instance, it is possible that they were created by adding the \texttt{WHERE} and \texttt{GROUPBY} clauses to a base query \texttt{SELECT a FROM T}.

\stitle{Special Cases: } Consider changes affecting the constants $5$ and $10$ in the following two queries:
{\small\begin{verbatim}
      SELECT b FROM T WHERE a > 5 AND a < 10
      SELECT b FROM T WHERE c > 5 AND a=10 \end{verbatim}
}
The first query involves filtering the values based on an interval, a natural fit for a range slider widget. The same widget would not apply for the second query. In order to distinguish between these two cases, \sys needs additional semantics to specify that the constants that change should be part of $>$ and $<$ inequality expressions that share the same attribute.

For these reasons, it is desirable to filter \difftable to a subset of query changes that are meaningful, as defined by the application and developer needs.  Note that \sys can be bootstrapped with a set of \lang statements for a given language so that, by default, it generates reasonable interfaces without any user intervention, and additional \lang statements can be added if desired.  It is an open area of investigation whether learning-based approaches can replace the need for manual \lang statements.

\subsection{\lang}
\lang is a domain specific language for users to easily specify {\it where} and {\it how} queries change.  A \lang statement $s$ is evaluated over a pair of queries $s(p_1, p_2)$ and returns an output table, or $\emptyset$ if it does not match. It is equivalent to filtering \difftable$_{\{p_1, p_2\}}$ by its $\pi$ attribute and transforming its subtrees $\tau_1$, $\tau_2$.  Although we will introduce simple tree traversal syntax that serves our purposes, more powerful nested SQL syntax such as SQL++~\cite{ong2014sqlpp} could be adopted in the future.

\lang{} statement comprises a From clause, a Where clause and a Match clause, organized as follows:

{\small
\begin{verbatim}
    	  FROM <path expression> AS <table name>, ...
    	[WHERE <boolean expression>]
    	 MATCH <stmt name>[(<table name>)]\end{verbatim}}

\stitle{From Clause: } The \texttt{FROM} clause is used to both define the query scope within which \sys searches for structural differences, and to transform the subtrees $\tau_1$ and $\tau_2$ in \difftable.  The path expression $path$ is composed of operators to specify ancestor \texttt{//} and child \texttt{/} relationships between node types; \texttt{*} denotes a wildcard node.

For instance \texttt{*//*} matches all possible paths, \texttt{a//*} matches any path containing node type \texttt{a}, \texttt{/a//*} matches paths whose root node is \texttt{a}, while \texttt{a/b} matches paths that contain \texttt{a} with direct child \texttt{b} that is also a leaf node.  If there are multiple matching nodes, \texttt{[i]} can be used to specify a specific child: \texttt{a/*[1]} selects the second child of \texttt{a}, while \texttt{a/b[1]} selects the second \texttt{b} child of \texttt{a}.

Since $path$ is always matched against $\text{\difftable}.\pi$, it will always match an ancestor of the subtrees $\tau_1$ and $\tau_2$; it is also used to specify the ancestor subtree to return.  The \texttt{FROM} clause returns the deepest subtree that matches $path$ and contains the subtree; the trailing \texttt{//*} specifies that the subtrees in \difftable should not be transformed.  In short, the \texttt{FROM} clause is equivalent to binding a range variable to the following SQL statement:
{\small
\begin{lstlisting}[
	mathescape,
	columns=fullflexible,
	basicstyle=\ttfamily\selectfont,
]
  SELECT id, pid1, pid2, extract($\pi$, path),
          ancestor($\tau_1$, path), ancestor($\tau_2$, path)
    FROM diffs
   WHERE matches($\pi$, path)
\end{lstlisting}
}

\stitle{Where Clause: } The \texttt{WHERE} clause is a boolean expression over the variables defined in the \texttt{FROM} clause. In addition to classic SQL expressions, path operators can be used to manipulate the subtrees $\tau_i$. Ellipsis notation (\texttt{..}) denotes the parent node, and the single dot (\texttt{.}) is used to access node attributes.  For instance, the following identifies tree differences within equality predicates:
{\small
\begin{lstlisting}[
	mathescape,
	columns=fullflexible,
	basicstyle=\ttfamily\selectfont,
]
    FROM Where/BiExpr AS T
   WHERE T.$\tau_1$.op = '=' AND T.$\tau_2$.op = '=' AND
          T.$\tau_1$/*[0].name = 'cty' AND T.$\tau_2$/*[0].name = 'cty'
\end{lstlisting}
}
The $\tau$ attribute can be used as shorthand for expressions over both subtrees.  Thus, the following is equivalent to the above example:
{\small
\begin{lstlisting}[
	mathescape,
	columns=fullflexible,
	basicstyle=\ttfamily\selectfont,
]
    FROM Where/BiExpr AS T
   WHERE T.$\tau$.op = '=' AND T.$\tau$/*[0].name = 'cty'
\end{lstlisting}
}
The following example checks for insertions in the query's project clause:
{\small
\begin{lstlisting}[
	mathescape,
	columns=fullflexible,
	basicstyle=\ttfamily\selectfont,
]
    FROM Project/ProjClause AS T
   WHERE T.$\tau_1$ is null AND T.$\tau_2$ is not null
\end{lstlisting}
}

\stitle{Match Clause: } This clause is used to name the \lang statement so that successful matches can be used as labeled edges in the interaction graph.  In addition, the statement returns one of the range variables so that it is accessible in for the interface generation step.  The returned range variable is augmented with a \texttt{name} attribute containing the statement's name.



\subsection{Executing and Writing \lang}
\label{sec:tool}

\lang statements are translated into SQL queries and executed over partitions of \difftable defined by a pair $(pid_1, pid_2)$.   The results over the partitions are unioned into a single table that we call \diffspil.  Section~\ref{s:opt} describes techniques that use \lang to reduce the tree matching and interface generation costs.

\lang is intended for developers, and crafting \lang statements manually may be difficult for users with no experience with abstract syntax trees. To facilitate this process, we created a tool to detect the most common differences between pairs of trees and let the users chose the ones that interest them. The tool operates as follows. The users specify a range of possible transformations, by setting a number of allowable pairwise differences and a number of nodes that may differ. The tool takes a sample from the log, compares all pairs of trees in that sample and reports those that match the specified criteria. To report the differences, it unparses the trees and highlights the substrings that vary. For any transformation, the tool can create a \lang statement that checks for differences in the subtrees where it detected variations. To help users refine their search, the tool excludes all the changes that match an existing \lang statement from subsequent searches.
\section{Interface Mapper}\label{s:interface}


The output of the interaction mining step is a table \diffspil representing subtree differences between pairs of queries in the query log.  We map this table into an interaction graph $G = (P_{log}, E)$  where each query is a vertex and each record is a directed edge $e = (p_i, p_j, l_e)$ that is labeled with a description of the interaction.  The graph is a multi-graph because each pair of queries can be connected by multiple transformations.  Our goal is to generate a set of interfaces $\mathbb{I}$ that can express the queries in this graph.  Doing so involves three challenges: 1) identifying candidate widgets for each edge in the graph, 2) extracting domain and template functions to instantiate those widgets, and 3) mapping subsets of the interaction graph to widgets in an interface. The remainder of this section describes how we tackle those problems.

\subsection{Preprocessing the Interaction Graph}
Recall that a widget $w = (t_\pi, \Omega_w, f_w)$ is instantiated with an interaction $t_\pi$, a domain $\Omega_w$, and a template function $f_w$ that maps elements in the domain to a subtree.  We now describe how to extract template functions and domains from \diffspil.  At a high level, we extract parameterized tree templates from the subtrees in the \diffspil table, and use the parameter values to construct the domains.  We will also use these reults to label the edges in the interaction graph.

\stitle{Template Functions: }  We use the subtrees in \diffspilns$.\tau_2$ to extract template functions.    To do so, we replace the $n$ primitive values in a subtree $\tau$ with $n$ parameters to create a parameterized subtree template $\tau^p$; the table $\mathbb{T} = (\tau^p, v_\mathbb{T})$ is modeled as its template $\tau^p$ along with its parameter values $v_\mathbb{T} = (v_\mathbb{T}^{i})_{i \in [1,n]}$.   We then group the trees by their templates, and for each group $g_{\tau^p}$, collect the parameter values for all trees in the group.  We then keep the indices $K \subseteq [1,n]$ of the parameters that have at least one change, and result in a per-group table $\mathbb{V}_{\tau^p} = \{(v_\tau^k | k \in K) | \tau \in g_{\tau^p} \}$.  This table represents the parameter values that vary across the same parameterized subtrees.

Each widget $w_i^\theta$ has a $k$-dimensional domain.  For instance a range slider has domain $(v_1, v_2) \in \mathbb{R}^2$.  A parameterized subtree can be mapped to a widget $w$ if there exists a bijection between attributes in its values in table $\mathbb{V}_{\tau^p}$ and dimensions in $w_i^\theta$'s domain, such that each binding is within $\Omega_\theta$.  If so, the template function $f_{w_i^\theta}$ is simply the inverse bijection from the widget's tuple state to the attributes in the values table that are then bound to the parameters in the template subtree.

\stitle{Labeled Edges: }  Each edge $e$ in the interaction graph represents a record $r$ in \diffspil.  We label the edge $l_e = (r.\pi, r.\tau_2^p)$ with its path $r.\pi$ and subtree template $r.\tau_2^p$.   We also annotate the edge with its parameter value $v_{r.\tau_2}$.   Finally, an edge could be mapped to multiple widgets---for instance, selecting from a set of options can be expressed by a dropdown and textbox---and we call these the edge's {\it candidate widgets} $W_e$.

\subsection{Widget Mapping}
The next step is to build a set of interfaces $\mathbb{I}$ that can express all the queries in the log.  To do so, we will first define the closure of an interface $I_{closure}$, and its cost $C_I$ in terms of the interaction graph, and show that the interface generation problem is NP-hard by a reduction from the {\it set cover} problem.   We then describe a graph contraction heuristic along with optimizations to speed up the process.  To simplify the description, we will first assume that each edge has a single candidate widget, each transformation is a scalar, and each \lang statement generates at most one edge between any pair of queries\footnote{\small{A \lang statement can generate an output table with multiple records, say due to multiple numbers changing, and each record corresponds to a transformation.}}.  We will then relax these restrictions.

An interface $I = (p^I_0, W^I)$ consists of an initial query $p^I$ and a set of widgets.  We define its closure as the set of reachable queries; a query $p_i$ is reachable by $I$ if there exists some path from $p^I_0$ to $p_i$ that consists of edges expressible by some widget in $W^I$.  An edge $e$ is expressible by a widget $w$ if its path and templated subtree are the same as the widget's.

The domain of widget $w \in W^I$ is defined as the union of the annotated values of the edges that it expresses in $I_{closure}$.   This domain is used to compute its cost $C_w(\Omega_w)$, and subsequently the cost of the set of interface $\mathbb{I}$.

\stitle{NP-Hardness: } We now sketch the reduction from set cover to the interface generation problem.
\begin{proof}
Given a universe of items $\mathbb{U} = \{u_1,\cdots,u_n\}$ and a set of $m$ subsets that covers $\mathbb{U}$, $\mathbb{S} = \{ S_i \subseteq \mathbb{U} | i \in [1, m]\}$, set cover identifies the {\it minimal} set of subsets $\mathbb{S}^* \subseteq \mathbb{S}$ such that $\mathbb{U} = \cup_{S \in \mathbb{S}^*} S$.

We can construct an interaction graph $G = (\mathbb{U}, E)$ where each subset $S_i$ forms a clique $\{ (u_i, u_j) | u_i, u_j \in S_i \} \subseteq E$, whose edges are labeled with the subset's id $l_e = S_i$.  The edges in $S_i$'s clique are expressible by a unique candidate widget $w_i$, and the cost of a widget is $0$ if its domain is empty, and $1$ otherwise.  Adding a widget $w_i$ to an interface adds all elements in $S_i$ into the closure, and the set of widgets in the resulting set of interfaces $\mathbb{I}$ forms the subset sum solution.
\end{proof}

\stitle{Simple Heuristic Solution: }  We present a greedy heuristic to solve the interface generation problem.  We initialize the solution $\mathbb{I}_0 = \{ I_i | i \in [1, |P_{log}|]\}$ by assigning an interface $I_i = (p_i, \{\})$ for each query $p_i \in P_{log}$.  We then greedily merge pairs of interfaces until the total cost of the interfaces does not further decrease.

A pair of interfaces $(I_i, I_j)$ are merge candidates if there exists zero or more edges that connect a query $p_i \in {I_i}_{closure}$ to a query $p_j \in {I_j}_{closure}$.  Let edge $e_{ij}$ be used to merge the interfaces.   The resulting merged interface $I_{ij} = (p^{I_i}_0, W^{I_{ij}})$ uses the initial query from $I_i$, and combines the widgets from both interfaces, along with the (single) candidate widget for the edge $e_{ij}$: $W^{I_{ij}} = W^{I_i} \cup W^{I_j} \cup W_{e_{ij}}$.   $W^{I_{ij}}$ can then be reduced by merging widgets that represent the same transformations: two widgets $w_a,w_b \in W^{I_{ij}}$ with the same paths and feature functions can be merged into a single widget $w_{ab}$ with domain $\Omega_{w_a} \cup \Omega_{w_b}$.

For each iteration $k$, we identify the pair of interfaces that, if merged, will most reduce the total cost:
\begin{align*}
(I_i^*, I_j^*) &= \argmax_{(I_i, I_j) \in \mathbb{I}_k\times\mathbb{I}_k} C_{I_i} + C_{I_j} - C_{I_{ji}}\\
\mathbb{I}_{k+1} &= (\mathbb{I}_k - \{I_i^*, I_j^*\}) \cup \{I_{ij}^*\}
\end{align*}

\begin{figure}[thb]
  \centering
  \includegraphics[width=.9\columnwidth]{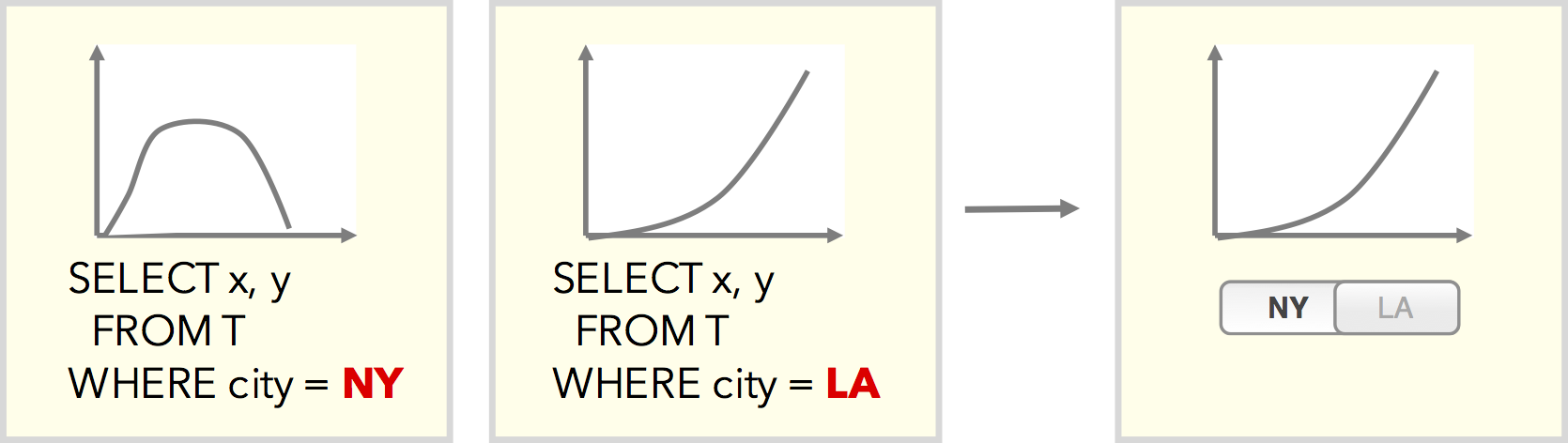}
  \caption{Merging two interfaces.}
  \label{fig:merging}
\end{figure}

\begin{example}
Figure~\ref{fig:merging} illustrates an example merge. The top two interfaces are initialized with their respective queries.  The queries differ in the constant in the equality predicate and there is a corresponding edge between the two queries. The interfaces are merged by mapping the edge to a toggle widget that picks between $NY$ and $LA$.  Note that the cost function  for the toggle widget will be high if the domain does not have exactly two values; if there are more queries with different $city$ values in the predicate, then other widgets such as a dropdown will have a lower cost and be chosen.
\end{example}

\stitle{Multiple Candidate Widgets:}
In practice, a given edge $e$ can have multiple candidate widgets $W_e$.  Rather than binding an edge to a specific widget in the initial interface set, we propagate all candidates throughout the heuristic solution.  To do so, we need to define how two candidate widgets are merged, how their costs are estimated.  Once we have finished the interface merging process, we then select the lowest cost interfaces from each candidate set. 

Merging candidate sets $W_{e_1}$ and $W_{e_2}$ means computing the union of the two sets $W_{e_1,e_2} = W_{e_1} \cup W_{e_2}$ and performing the domain merging procedure described above. The cost of a candidate set $C_{W_e} = min(C_w | w \in W_e)$ is defined by the minimum cost widget in the set.


\stitle{Multiple Edges:}
A given \diffspil table for queries $p_i, p_j$ can contain $n>1$ records, each representing an edge with a different path between the pair of queries in the graph.  However, in order to fully transform $p_i$ into $p_j$, we must apply the transformations for {\it all} the edges.   To account for this, we model these edges $e_1,\cdots,e_n$ as a single ``super-edge'' $e_{1\cdots n}$  whose candidate widget set is the cross product of each edge's candidate set $W_{e_{1\cdots n}} = \varprod_{i=0}^n W_{e_i}$.

\stitle{Collection-based Changes}
Most languages support expressions that represent collections; for instance the SQL \texttt{FROM} clause is a set of range variable definitions, and the \texttt{GROUPBY} clause is a list of grouping expressions. Yet, the model described in Section~\ref{sec:diff} does not account from interactions that manipulate sets.  For instance, the output of a \lang statement that looks for numerical differences over the program fragments \texttt{[1,2,3]} and \texttt{[1,4]} would output a table with the following $(\tau_1, \tau_2)$ pairs: $(2\rightarrow4), (3\rightarrow null)$.  Naively, each pair would be modeled as a separate widget. In many cases, it would be preferable to map the pairs to a multi-selection widget such as a checkbox list that can express collections.

To automatically translate such \diffspil tables into a single collection-based interaction, we use a procedure similar to extracting template functions.  We first collect the set of all subtrees in \diffspil: $\mathrm{T} = \cup_{r \in \diffspil} \{ r.\tau_1, r.\tau_2 \} $.  For each $\tau \in \mathrm{T}$, we look for the subtree $\phi_\tau$ rooted at its closest strict ancestor node that is annotated as a list node type. If no such subtree exists one or more subtrees, we stop.  Otherwise, we then replace $\tau$ in $\phi_\tau$ with a parameter variable to create a templated ancestor subtree $\phi^p_\tau$.  If all templated ancestors are identical, then that suggests that the subtrees are elements of a collection, and we map the entire \diffspil table to a single collection-interaction edge whose candidate widgets are collection-based widgets such as checkbox or multi-select.

\subsection{Generating Interfaces}
Once we have identified the optimal set of interfaces $\mathbb{I}^*$, we select the lowest cost widget from each of the candidate sets based on the final widget domains.  At this point it is possible to run a standard interface layout algorithm~\cite{sears1993layout}, and then render each interface as a tab in a web application.  We render the query output using the developer provided $exec()$ function.  In our implementation for SQL query logs, we position the widgets manually and we use a simple visualization generator similar to ShowMe~\cite{mackinlay2007show} or APT~\cite{mackinlay1986automating} if the number of attributes in the query output is small, and otherwise render a table.



\section{Optimization}\label{s:opt}

A naive implementation of \sys first materializes \difftable by computing tree alignments between all pairs of ASTs in the query log, filtering \difftable using the \lang statements, transforming the results into an interaction graph, and performing the graph contraction procedure to derive the interfaces.  However, the cost of these steps, in particular tree alignment and graph contraction, can be considerable for even thousands of queries in the log. Fortunately, we may exploit three properties of the problem to reduce the number of pairwise comparisons: the transitivity of transformations,the existence of templates and the fact that all transformations may not be relevant.

\subsection{Transitive \lang Cliques}

\label{transcliques}
A \lang statement $s$ is transitive if matches between $(p_1, p_2)$ and $(p_2,p_3)$ implies a match between $(p_1,p_3)$:
{
  $$s(p_1, p_2) \ne \emptyset \wedge s(p_2, p_3) \ne \emptyset \rightarrow s(p_1, p_3) \ne \emptyset$$
}
If $s$ is transitive, then the set of programs $C$ that it matches forms a clique, and a new program $p$ need only compare with an arbitrary program $p_c\in C$ to check if matches with all members of the set.  Algorithm~\ref{a:clique} uses this observation to efficiently evaluate transitive \lang statements directly on the program log.
{\small
\begin{algorithm}
 \caption{Clique detection for transitive \lang statements.}
 \label{a:clique}
 \KwData{\lang statement: s, Programs: P}
 \KwResult{C}
 initialize C = $\{\}$\;
 \While{$p \in P$}{
  matched = false\;
  \While{$c \in C$}{
    \If{$\exists_{p_c \in C} s(p, p_c) \ne \emptyset$}{
     c.add(p)\;
     matched = true\;
     }
  }
  \If{$\neg$\textrm{matched}}
  {
    C.add($\{p\}$);
  }
 }
\end{algorithm}
}

We use a simple heuristic that identifies transitive statements in the \lang statements we have developed.  We check that the \texttt{WHERE} clause only contains transitive logical expressions (e.g., $=, \ne$).  We leave richer analysis techniques as future work.

A welcome side effect of this method is that it allows us to \emph{compress} the interaction graph. Indeed, this graph is often unpractical because it is very dense. Typically, it contains $\mathcal{O}(N^2)$ edges for $N$ queries, a consequence of the \lang's statements transitivity. Thanks to Algorithm~\ref{a:clique}, we can store it as a set of \emph{cliques} rather than a set of \emph{edges}, and lower the storage cost by an order of magnitude. We will show in Section~\ref{sec:atscale} that this strategy let us process large query logs in main memory.

\subsection{Program Templates}\label{s:templates}

We observe that query logs often contain cliques due to queries that have identical parse structures (i.e., {\it templates}) but different values in the literals.  For instance, queries emitted by varying a distance threshold or function parameter are identical everywhere except for a single value.  We term these cliques {\it templated cliques}.
To detect such cliques, our procedure is similar to finding template functions in Section~\ref{s:interface}: we replace the literals in the query ASTs (e.g., all node attributes are primitive values) with unnamed variables, hash the resulting templates and group by the hash values. Thus, we represent a group of similar ASTs by a template followed by a list of literals.  The rest of the system performs tree alignment and \lang evaluation over templated cliques rather than individual queries.  For each templated clique, we index the paths to each variable and probe each clique with the path expressions in each \lang statement.  A matching statement can use the index to quickly identify the ASTs that change and evaluate those; a statement that does not match can skip the clique altogether.

Since these operations do not rely on user inputs (e.g., \lang statements), we can perform them offline, during a preprocessing step. Furthermore, the output can be reused as the developer adds new \lang statements and tunes the interface generation parameters.

\subsection{Restricting Program Pairs}

Since the interfaces generated by \sys are derived from differences between pairs of programs, we can restrict the pairs to compare for performance and personalization purposes.   For instance, we might only compare program pairs in sequence (e.g., $p_i$ with $p_{i+1}$), or filter the program log table \texttt{progs} (Section~\ref{s:model-progs}) by the user, timestamp or other metadata.  Modeling the program log as a relation provides the system with considerable flexibility in choosing the program pairs.

\section{Experiments}
\begin{figure*}[t!]
    \centering
    \includegraphics[width=.78\textwidth]{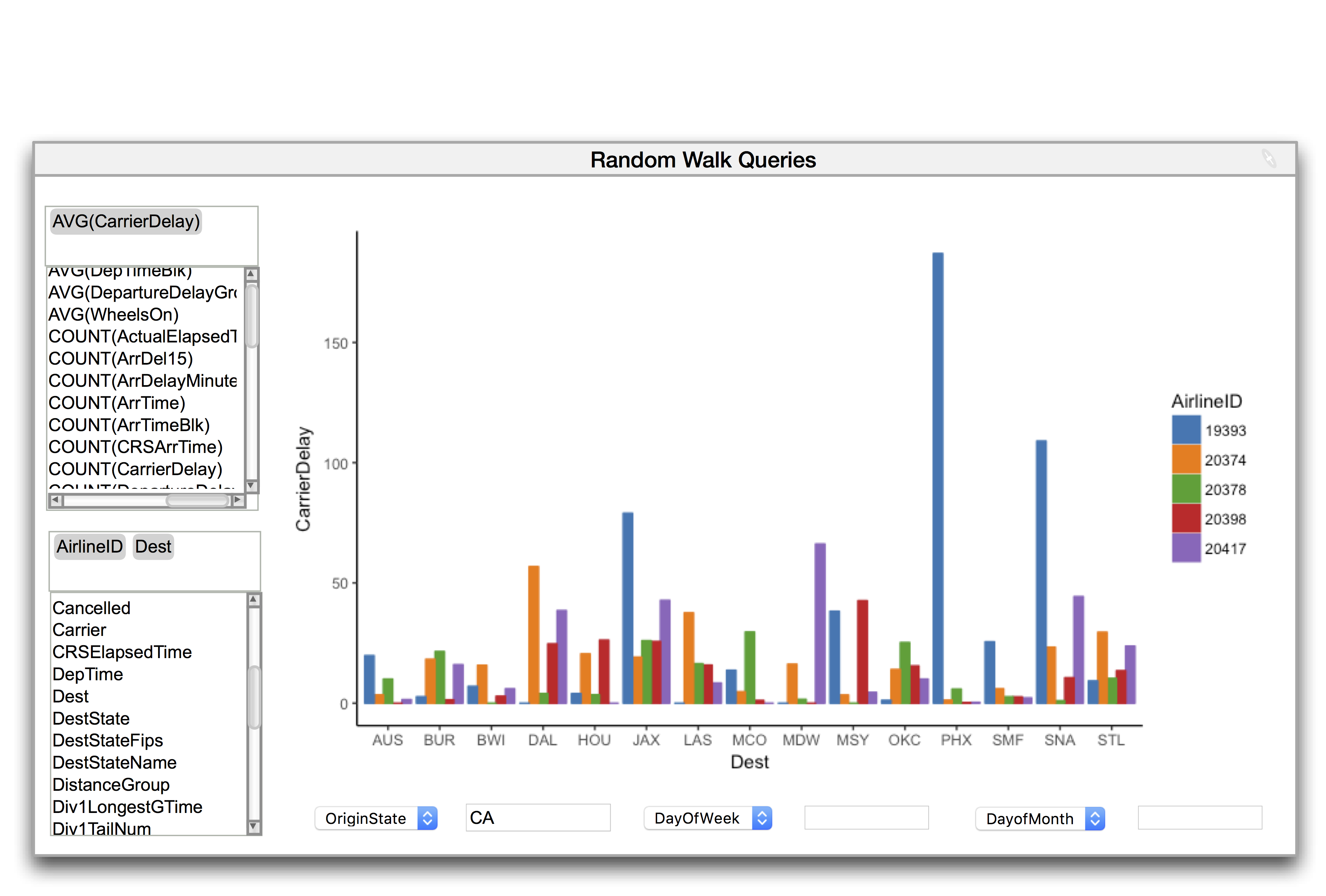}
    \caption{Interface generated from a log of random OLAP queries over the On-Time database. The widgets were created by \sys, we edited the layout and titles manually. Users choose dimensions and measures by dragging and dropping in the leftmost boxes. They create filters with the dropdown lists and text boxes at the bottom of the screen.}
    \label{fig:ui-randomwalk}
\end{figure*}
\begin{figure}[t!]
    \centering
    \includegraphics[width=\columnwidth]{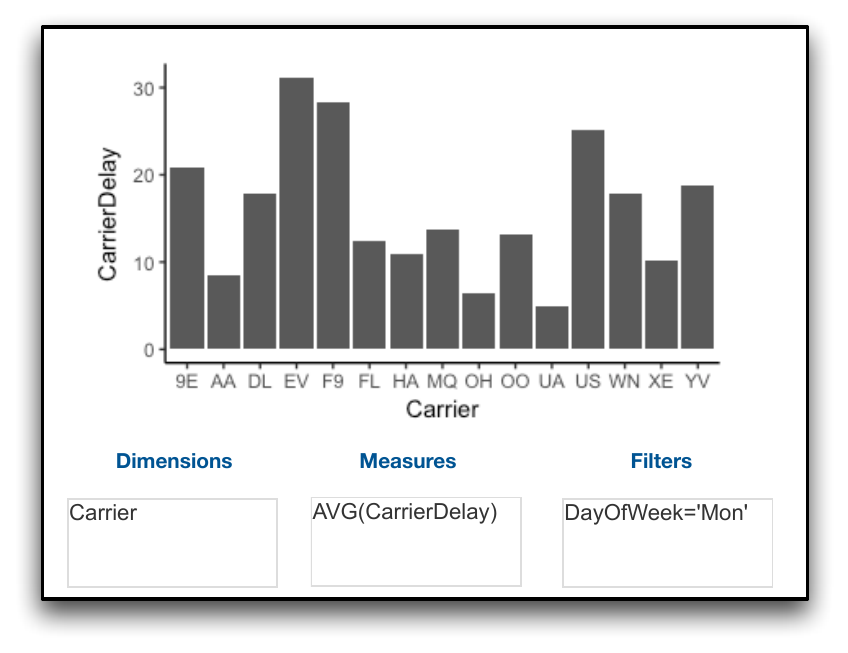}
    \caption{Interfaces generated for the random OLAP queries. Values simplicity.}
    \label{fig:ui-simple}
\end{figure}
\begin{figure}[t!]
    \centering
    \includegraphics[width=\columnwidth]{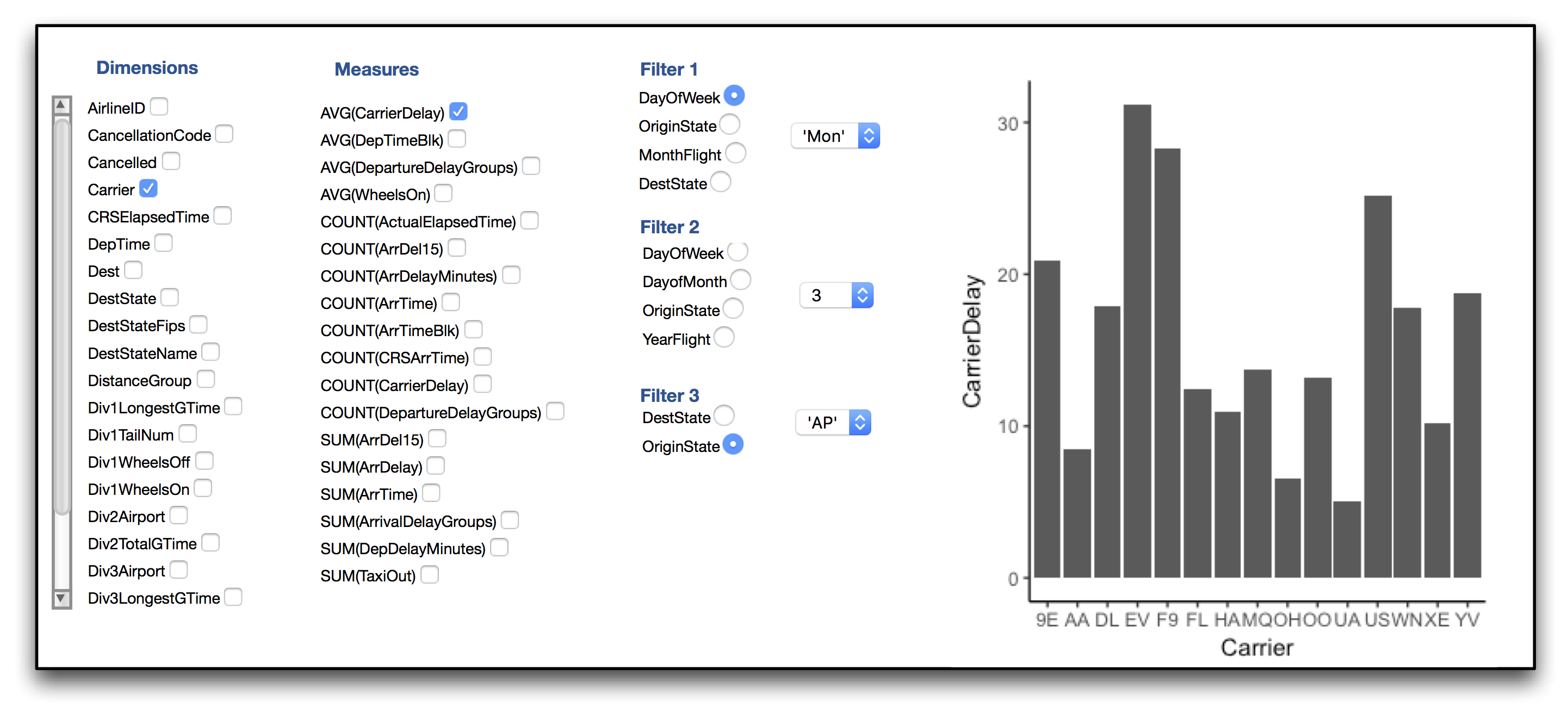}
    \caption{Interfaces generated for the random OLAP queries. Values directness.}
    \label{fig:ui-direct}
\end{figure}
\begin{figure*}[t!]
    \centering
    \begin{subfigure}[b]{0.25\textwidth}
        \includegraphics[height=1.8in]{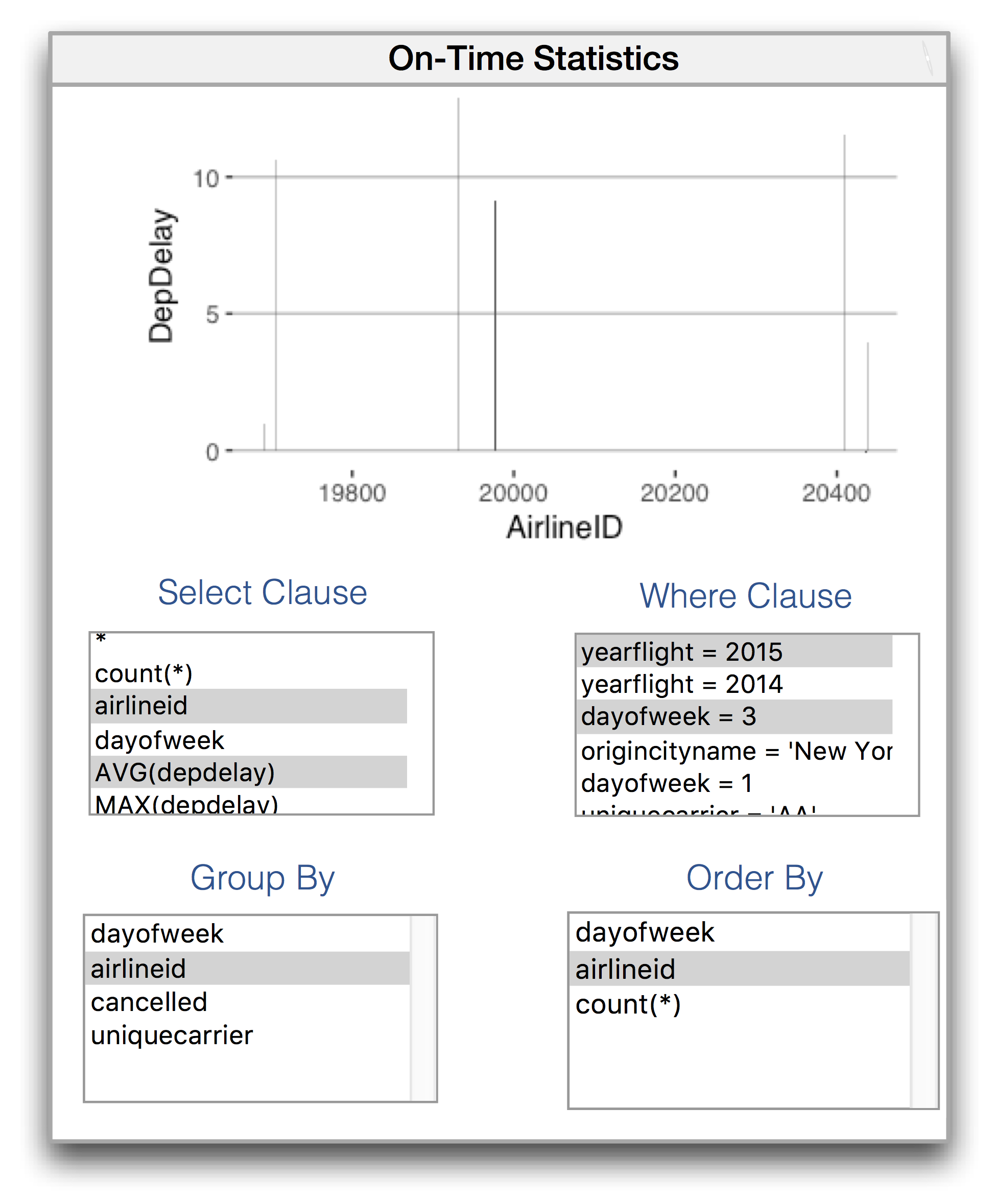}
        \caption{First interface.}
        \label{fig:ui-manual1}
    \end{subfigure}
    ~
  \begin{subfigure}[b]{0.25\textwidth}
        \includegraphics[height=1.8in]{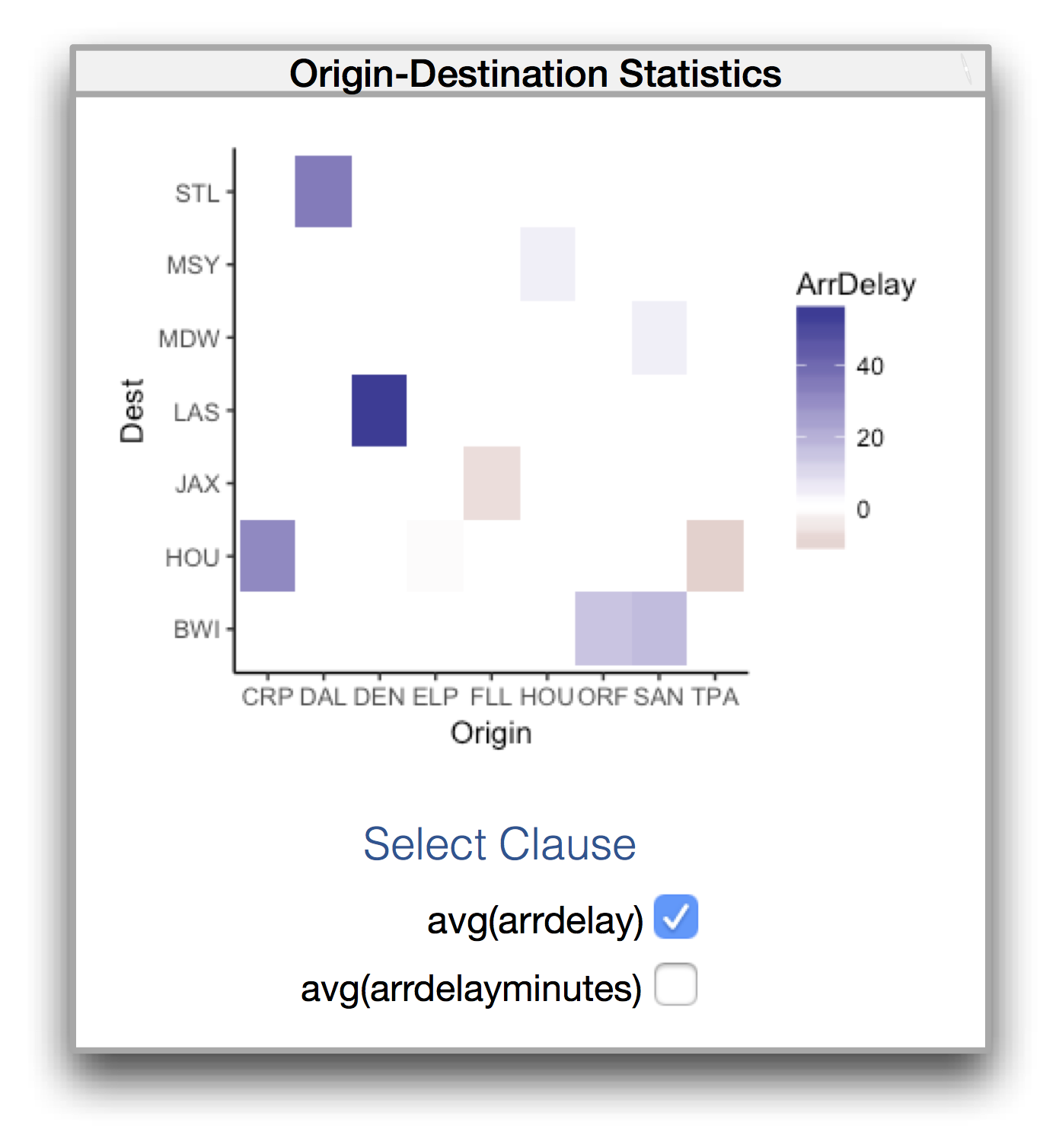}
        \caption{Second interface.}
        \label{fig:ui-manual2}
    \end{subfigure}
    ~
  \begin{subfigure}[b]{0.25\textwidth}
        \includegraphics[height=1.8in]{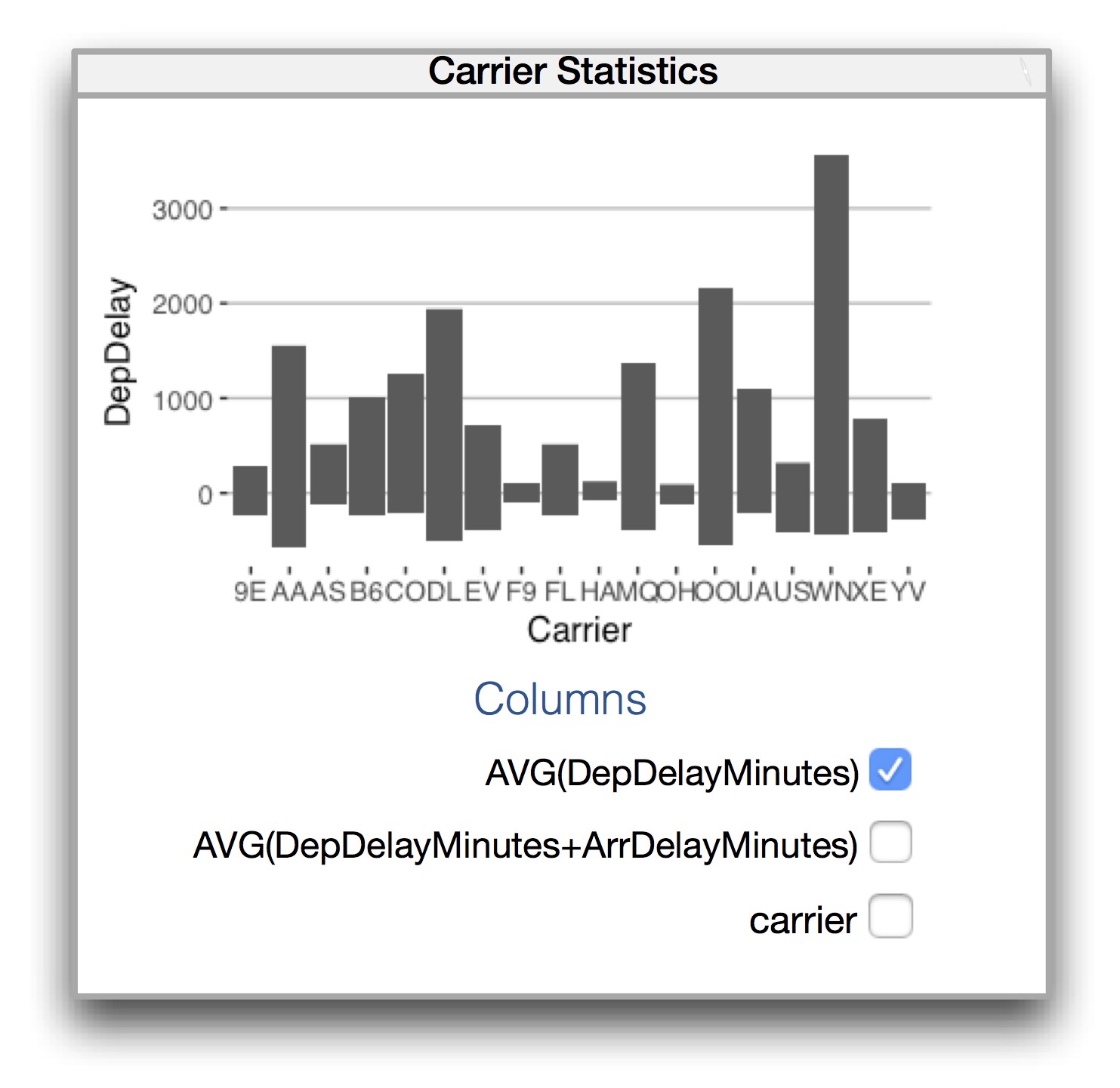}
        \caption{Third interface.}
        \label{fig:ui-manual3}
    \end{subfigure}
    \caption{\small Top 3 interfaces generated by \sys{} from a set of queries over the On-Time database written by students. The first window lets users compose simple \texttt{group by} queries, by selecting lines in the list boxes. The second one presents aggregates for each combination of origin and desintation airport. The last interface shows statistics for each carrier. Collectively, those interfaces cover 59\% of all the queries that we collected.}
      \label{fig:ui-manual}
\end{figure*}
We evaluate \sys using 5 query logs---4 SQL and 1 SPARQL log.  We used simulated query logs, logs from existing data systems, and logs generated through manual visual exploration, in order to study the system on clean, real-world, and ad-hoc types of logs types, respectively.   We seek to answer four questions: 1) can \sys{}'s interfaces express query logs? 2) is the runtime acceptable? 3) can the system support multiple languages?  and 4) how do users prefer the generated interfaces as compared with existing and user-designed interfaces?

Parsing and interaction mining are implemented in Java, and the widget mapping and rendering are implemented in Python, which generates HTML+Javascript interfaces.  We defined 12 widget types (e.g., dropdown, checkbox list, slider, range slider, textbox, multi-select) and manually created $exec()$ and $render()$ functions for SQL and SPARQL.   After generating the interfaces, we named and positioned the widgets for presentation purposes. We used a MacBook Pro with Intel Core i7 2.5 GHz CPU and 8GB RAM.

\subsection{Expressing Query Logs}
\label{sec:case_stud}

We first showcase the generated interface designs and their ability to express queries from two logs---synthetically generated OLAP queries, and those generated through manual exploration---over the On-Time Database\footnote{\small 521,000 rows, 91 cols. \url{https://www.transtats.bts.gov}}.

\stitle{Synthetic OLAP queries:} The aim of this experiment is to show that \sys{} can generate a simple Tableau-like interface from a standard OLAP query workload.
To simulate the exploration process, our generator explores the OLAP space by starting with a group by-aggregate query, and iteratively modifying a random clause (GROUPBY, WHERE, SELECT) at each step. To seed the process, we generate a random group by-aggregate query that follows the following format:
\begin{verse}
\texttt{SELECT dim1, \ldots, dimM,}\\
        \texttt{~~~~~~~agg1(meas1), \ldots, aggN(measN)}\\
\texttt{FROM Ontime}\\
\texttt{WHERE var1=val1 AND \ldots AND varP=valP}\\
\texttt{GROUP BY dim1, \ldots, dimM}\\
\end{verse}
The number of dimension, filters and measures is a sampled from a uniform distribution. We then perform random edits, one for each step. We present the possible edits in Table~\ref{tab:changes}. We wrote 7 PILang statements that correspond to structural and value changes that our query generator expressed.

\begin{table}
\caption{Possible modifications from the query generator.}
\begin{tabular}{ c l }
\hline
  Type &  Actions\\
  \hline
  Dimensions &  Add, Remove, Change\\
  Measures   & Add, Remove, Change col., Change agg.\\
  Filters    &  Add, Remove, Change col., Change val.\\
  \hline
\end{tabular}
\label{tab:changes}
\end{table}

Figure~\ref{fig:ui-randomwalk} presents the generated interface. The two drag-n-drop boxes on the left let users choose measures and dimensions to visualize. The bottom section of the interface provides three filters, each consisting of a drop-down list to select a column and a text field to specify a value. This interface can express 100\% of the queries in the log (i.e., its closure contains all the queries in the interaction graph). As Tableau, it lets users produce OLAP queries by dragging columns onto ``shelves'', however further work is needed to generate complex logic such as small multiples.

Figure~\ref{fig:ui-simple} presents an alternative UI, obtained from the same set of queries but with different parameters. In this case, we tuned the weight associated to the cost functions to obtain the most simple interface possible. We assigned a high cost to visual complexity and a low cost to user effort. As a result, the UI contains only three text boxes---one for the dimensions, one for the measures and one for the filters. The user must type the queries manually.

To generate Figure~\ref{fig:ui-direct}, we reversed the weighing scheme. We assigned a high cost to user efforts (e.g., number of keyboard interactions and clicks) and ignored visual complexity. In the resulting window, all the options are explicit: there is one tick box for each possible dimension, measure of filter column, and a dropdown lists for the filter values.

\stitle{Manual log:} We created an ad-hoc exploration query log by asking 12 students to perform 3 random (out of 12) tasks using the On Time dataset (e.g., ``how delayed are the flights to from AA?''), answer one free form question (``tell us something you found surprising'') and report their findings.   We logged all queries that were executed.  There are a total of $298$ statements, $148$ unique. We did no clean the log (e.g., dead-end analyses, erroneous queries) and simply report the top interfaces.  We used $15$ \lang statements writtend using the tool described in Section~\ref{sec:tool}.

\begin{figure}[h!]
    \centering
    \includegraphics[width=.7\columnwidth]{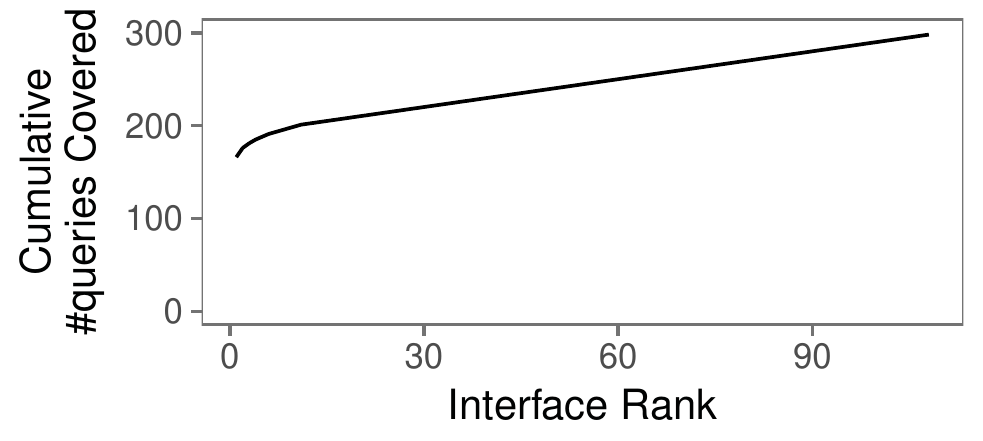}
  \caption{Coverage as more interfaces are added to $\mathbb{I}^*$ for manual log. Shows long tail of queries.}
    \label{fig:ui-manual-coverage}
\end{figure}
This query log contains far more variability than the synthetic dataset, and we consider it a ``hard'' case. Figure~\ref{fig:ui-manual-coverage} shows the total number of queries covered as the number of interfaces in the output $\mathbb{I}^*$ increases. We observe that the first interface covers $166$ ($55\%$) queries, and subsequent interfaces cover $<10$ queries each. This suggests that the interaction graph is sparse, which is reflected in our post-hoc analysis of the logs. 

Figure~\ref{fig:ui-manual} shows the top three interfaces. The left interface is the primary one that resembles a simplified Tableau: most students incrementally vary the select, where, groupby and orderby clauses.  The middle interface covers $10$ ($3\%$) and computes aggregate statistics for each flight origin; the right interface is representative of the long tail (covers $1-3$ queries). Although those three interfaces do not cover the whole log, they express the primary exploration structure using only 6 interaction components. 

\subsection{Performance and Languages}
\label{sec:atscale}

In this experiment, we evaluate \sys{}'s language support and scalability. To evaluate the first aspect, we run the pipeline on logs written in two different query languages. To
test the second, we measure its runtime for different optimizations. We show that \sys{} spends more than 90\% of its time in the interaction mining stage, and therefore we focus on this step.

We use two programs logs. The first one is the {\it SDSS} log~\cite{sdss}, which contains $125,603$ SQL queries ($112,847$ unique).  The second is a sample from the {\it British Museum's} Semantic Web Collection~\cite{britishmuseum}, which contains $110,677$ SPARQL queries ($38,933$ unique). We respectively used $16$ and $4$ \lang statements for SQL and SPARQL, which describe the more frequent transformations, detected both by manual inspection and by using the tool described in Section~\ref{sec:tool}. We compare four settings: no optimization, the clique-based optimization of Section~\ref{transcliques} (Clique), the program templates of Section~\ref{s:templates} (Template), and both optimizations. By default, the latter setting is enabled. Our main finding is that using both optimizations allows \sys to scale to logs that are two orders of magnitude larger than without any optimization, and thus it can process logs with 10,000s of queries in minutes.

\begin{figure}[h!]
    \centering
    \includegraphics[width=.8\columnwidth]{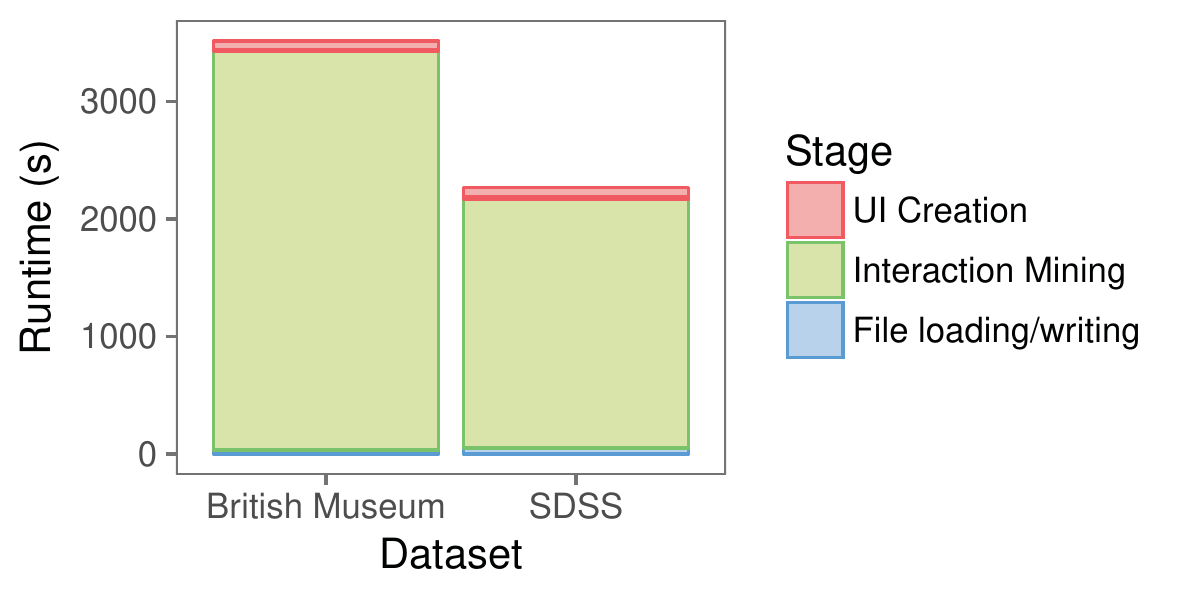}
    \caption{Breakdown of execution time for each dataset.}
    \label{fig:breakdown}
\end{figure}
\stitle{Cost Breakdown:} Figure~\ref{fig:breakdown} shows the overall cost breakdown. The Interaction Mining phase is by far the most time consuming. Because this cost largely dominates \sys{}'s runtime, the rest of this section focuses on it.

\begin{figure}[h!]
    \centering
    \includegraphics[width=\columnwidth]{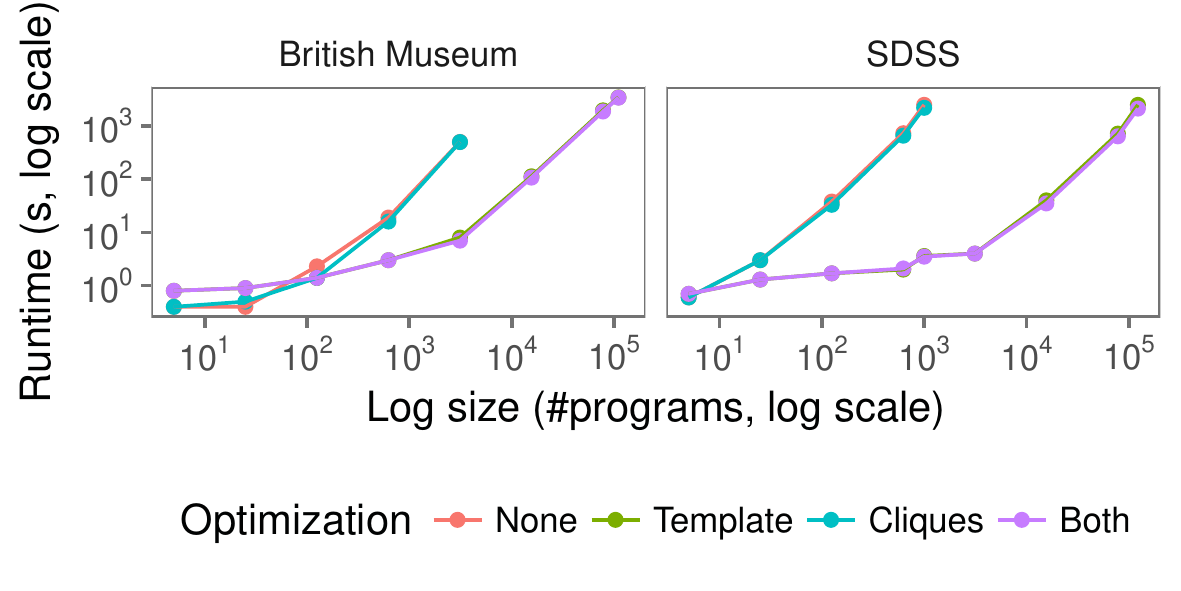}
    \caption{Runtime (secs) of interaction mining vs log size (log scale). The points for Template and Both overlap. We cap maximum runtime to $1$ hour.}
    \label{fig:time-both}
\end{figure}
\stitle{Scalability with the size of the program logs:} Figure~\ref{fig:time-both} shows that the runtime increase quadratically for both logs, even when the optimizations are enabled. This comes from the fact that \sys{} must align and compare $\mathcal{O}(N^2)$ pairs of programs to build the interaction graph, where $N$ is the number of programs. We ran a micro-benchmark and found that the cost of the comparisons is almost constant---they take in average $3.4\pm0.08ms$ for the SDSS log and $1.20\pm0.01$ms for British Museum log ($\pm$ are 95\% CIs). The cost comes from the high number of comparisons.

The optimizations do not reduce the $\mathcal{O}(N^2)$ worst-case complexity of the algorithm, but they allow the system to skip comparisons. In particular, the Template optimization incurs a runtime improvement of about two orders of magnitude  compared to No Optimization---up to 347x for the SDSS data set and 71x for the British Museum data.

\begin{figure}[h!]
    \centering
    \includegraphics[width=\columnwidth]{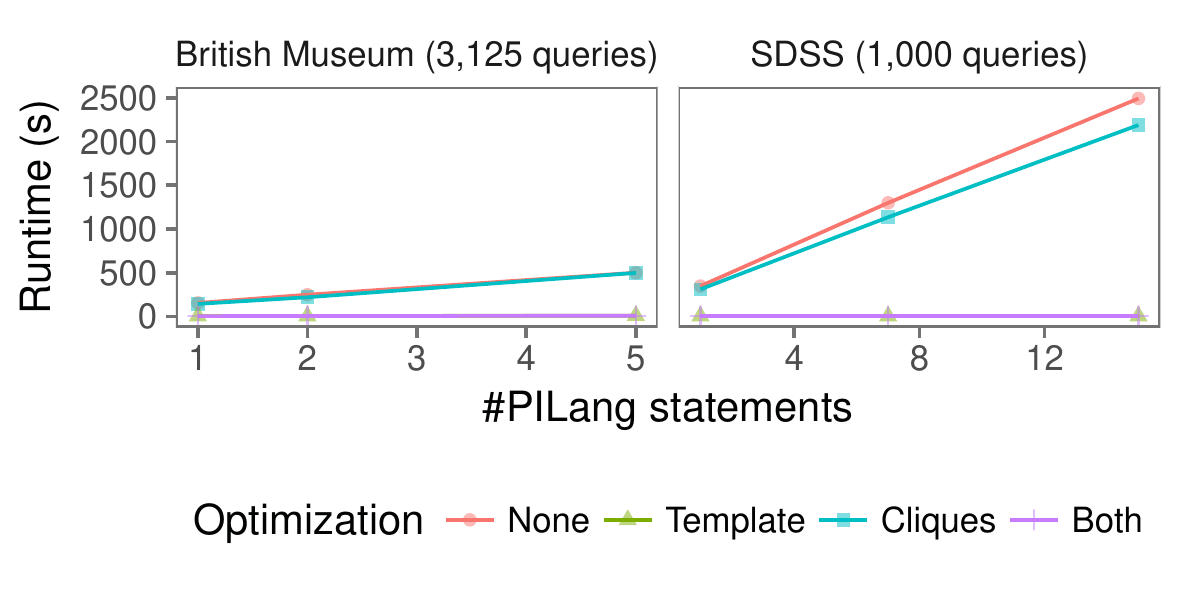}
    \caption{Runtime of interaction mining vs. number of PILang statements, for samples of each dataset. In both cases, the plots for Cliques and Both overlap.}
    \label{fig:pl-sdss}
\end{figure}
\stitle{Scalability with the number of \lang{} statements:} Figure~\ref{fig:pl-sdss} presents how the interaction miner's runtime changes when we vary the number of \lang statements. The number of queries is fixed; we used small sample sizes to enforce that all the versions of the algorithm reach completion within one hour. In both cases, we find that increasing the number of statements linearly increases the runtime. Here again, the Template optimization yields improvements of about two order of magnitudes.

\begin{figure}[h!]
    \centering
    \includegraphics[width=\columnwidth]{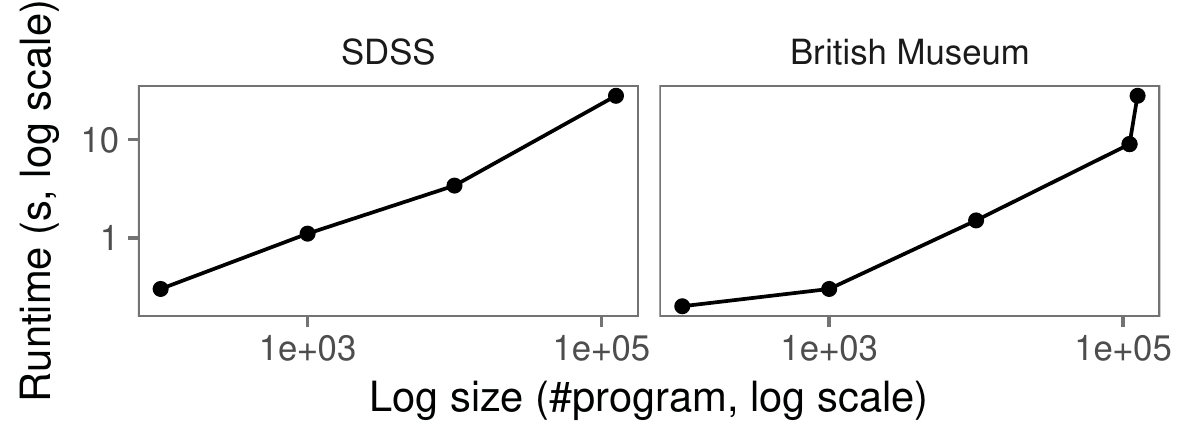}
    \caption{Templated clique extraction costs for each log.}
    \label{fig:prepro}
\end{figure}
\stitle{Preprocessing: } Figure~\ref{fig:prepro} shows the time required to extract the query templates (Section~\ref{s:templates}). The runtime increases quadratically with respect to the log size, but it runs within $30$s for both program logs. 

\begin{figure}[h!]
    \centering
    \includegraphics[width=.7\columnwidth]{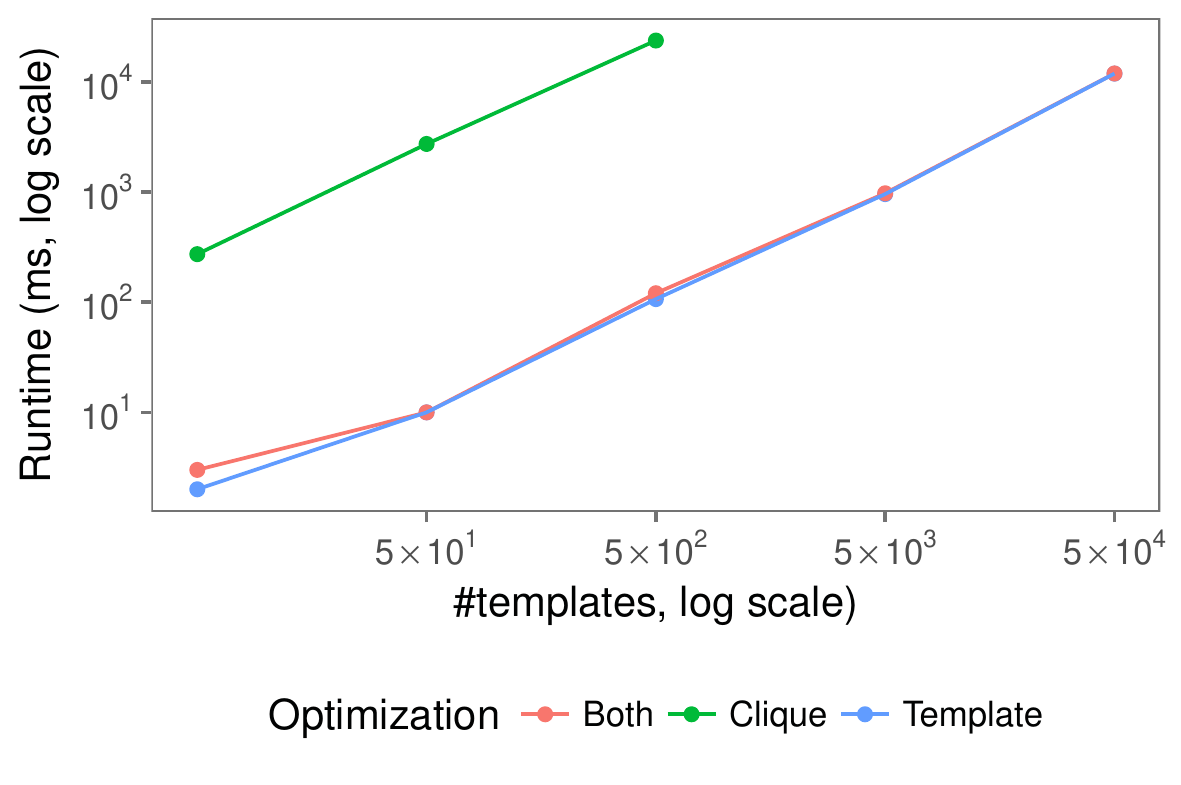}
    \caption{Runtime of interaction mining for a log of 50,000 queries, varying the number of templates.}
    \label{fig:ui-template}
\end{figure}
\stitle{Impact of the Structure on the Optimizations: }
In this set of experiments, we show how the optimizations' efficiency varies with the structure of the programs in the log. We generate random queries using templates; we fix the number of queries, vary the number of templates and measure how \sys{}'s runtime varies. We expect that the runtime decreases as the variability of the queries (i.e. the number of templates) decreases.

Figure~\ref{fig:ui-template} presents the results of the experiments. We observe that the runtime of interaction mining varies linearly with the number of templates, for a fixed number of queries. The effect is similar for both Clique and Template, though Clique is two orders magnitude slower than Template. This illustrates that the optimizations successfully exploit the structure in the log. The more structure the log contains, the faster interaction mining runs.

\begin{figure*}[t!]
    \centering
  \begin{subfigure}[b]{0.28\textwidth}
	    \includegraphics[width=\textwidth]{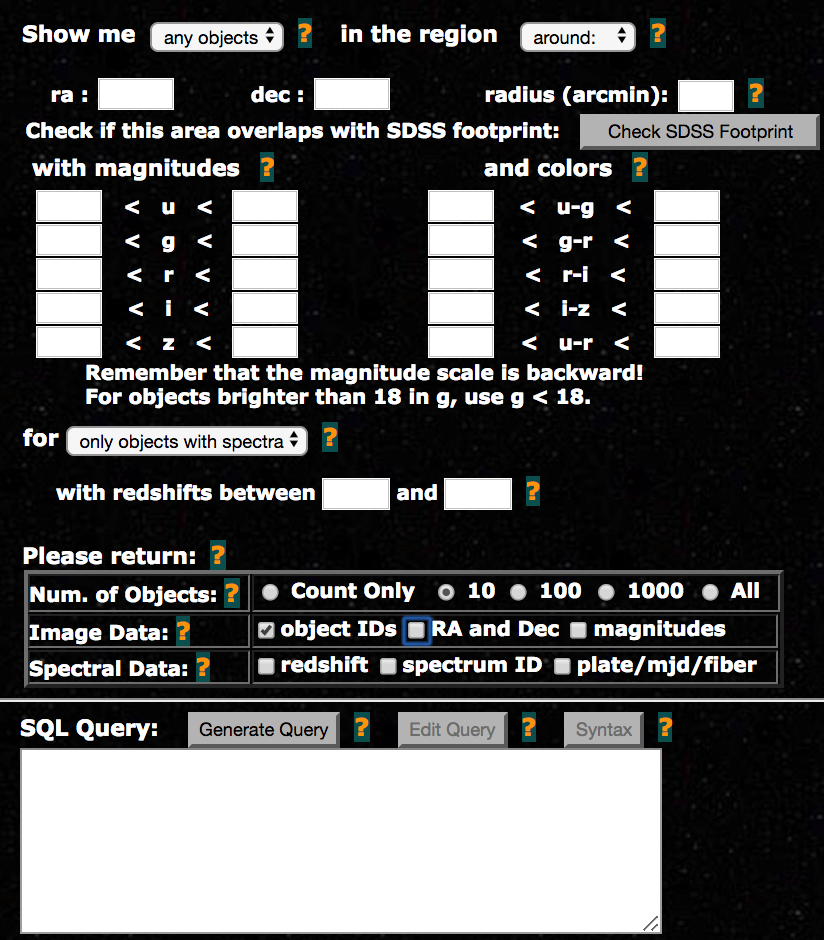}
	    \caption{Original SDSS Interface.}
	    \label{fig:ui-1}
	\end{subfigure}
	~
  \begin{subfigure}[b]{0.18\textwidth}
	    \includegraphics[width=\textwidth]{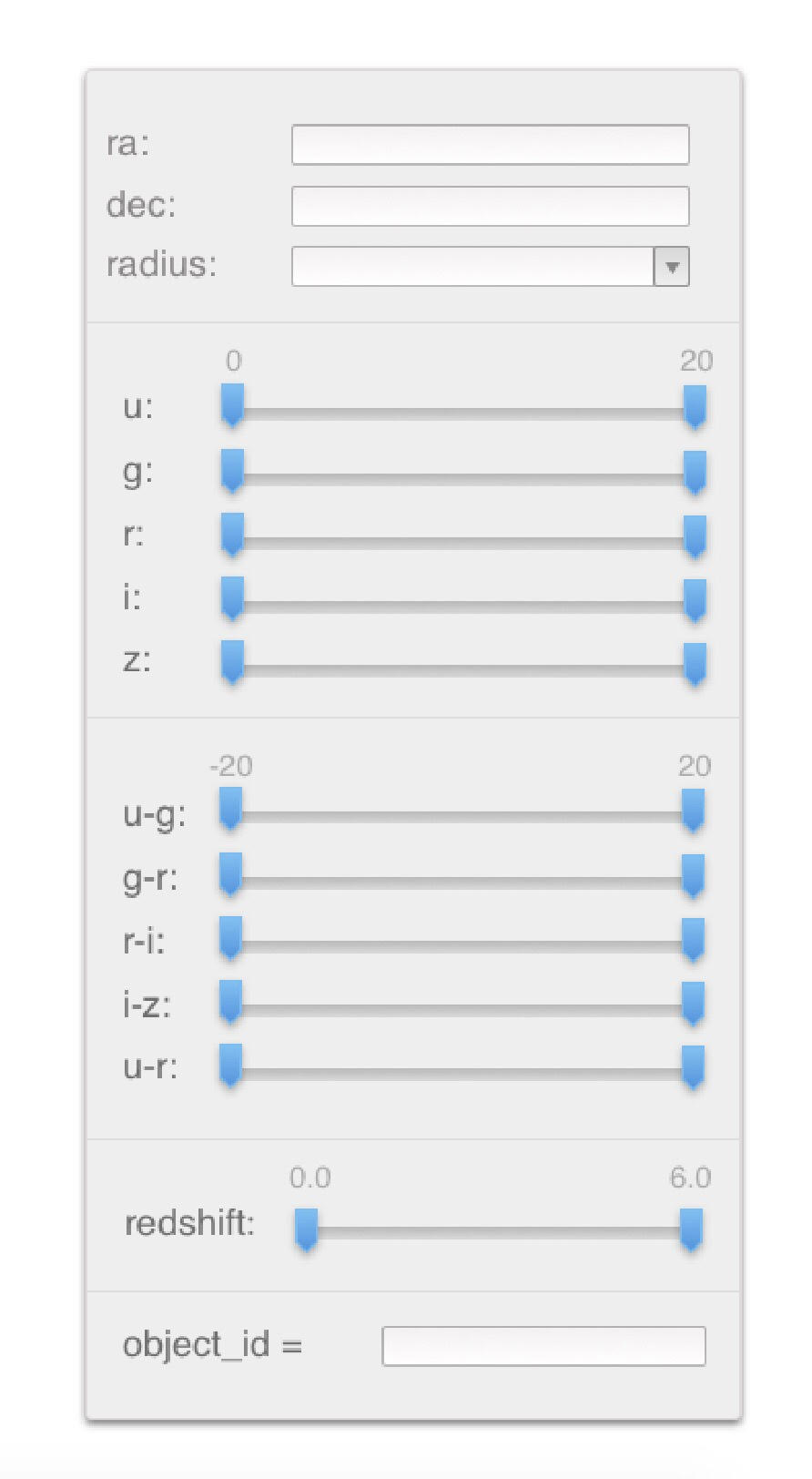}
	    \caption{Precision Interface}
	    \label{fig:ui-2}
    \end{subfigure}
	~
  \begin{subfigure}[b]{0.18\textwidth}
	    \includegraphics[width=\textwidth]{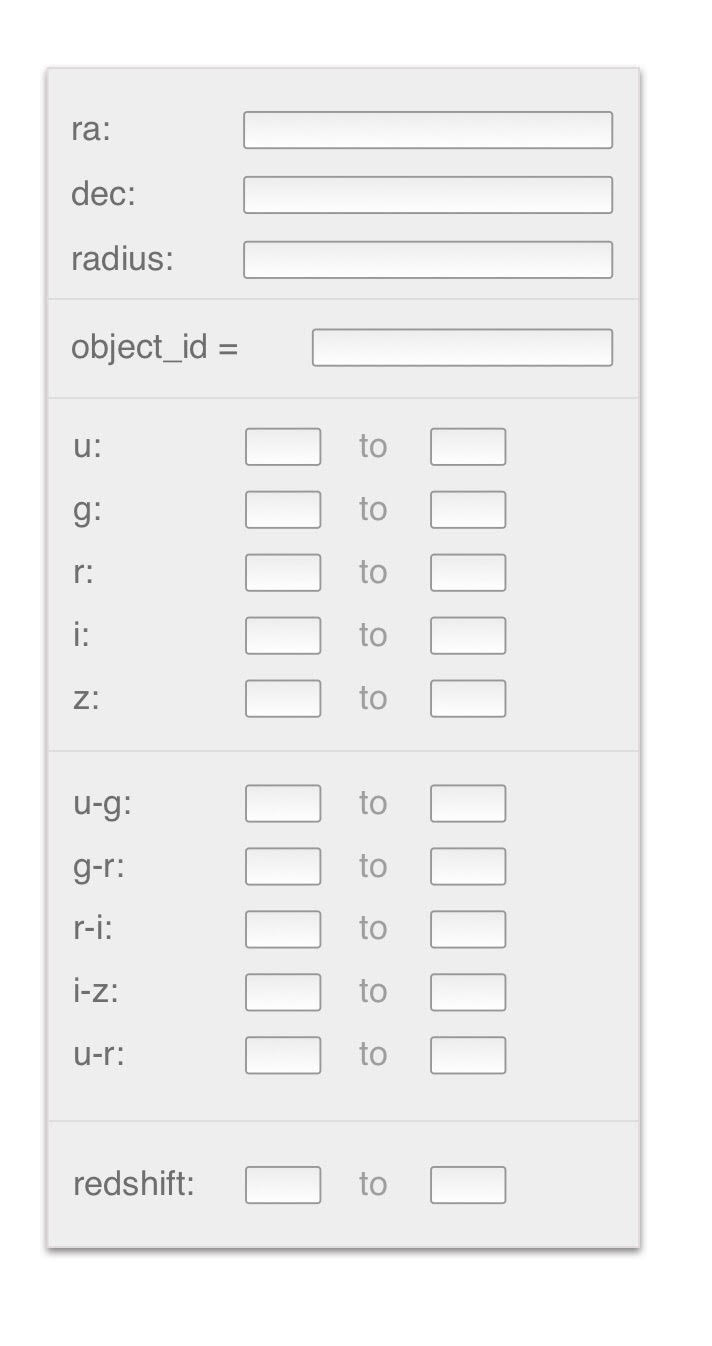}
	    \caption{Custom Interface 1}
	    \label{fig:ui-3}
  \end{subfigure}
    ~
  \begin{subfigure}[b]{0.28\textwidth}
	    \includegraphics[width=\textwidth]{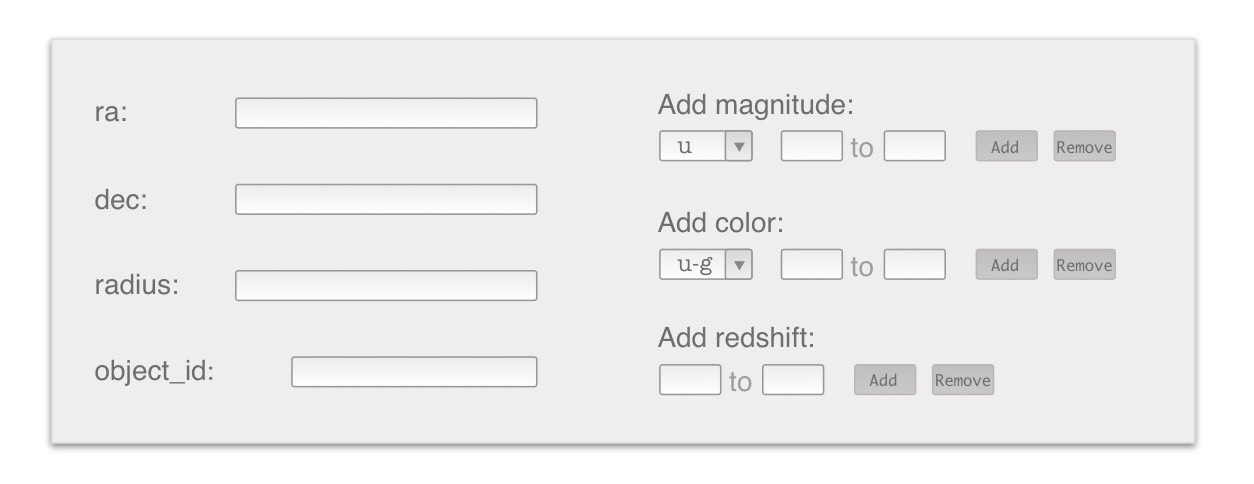}
	    \caption{Custom Interface 2}
	    \label{fig:ui-4}
  \end{subfigure}
  \caption{The original SDSS interface, the interface generated by \sys, and two manually designed interfaces.}
  \label{fig:allinterfaces}
\end{figure*}

\subsection{User Study}
We conducted users studie based on the SDSS query log, using the original Sky Server interface\footnote{\url{http://skyserver.sdss.org/dr14/en/tools/search/form/searchform.aspx}} for reference. We studied whether 1) the generated interfaces reduce the reponse time and analysis accuracy as compared to the existing interface and 2) the generated interfaces are competitive with the original one and handcrafted alternatives in terms of user preference. We recruited 40 CS graduate students for the study.

\stitle{Comparison With Existing Interface}
 Users were given 5 minutes to read the manual\footnote{\url{skyserver.sdss.org/dr9/en/tools/search/}} describing the 4 tasks supported by the existing SDSS interface (Figure~\ref{fig:ui-1}), and interact with the interface.  To avoid learning effects, we randomly split the users into two groups which were asked to complete the 4 tasks using different interfaces. The first group used the existing interface, while the second used \sys (Figure~\ref{fig:ui-2}). We recorded the analysis time and result accuracy for each task.

\begin{figure}[h!]
    \centering
    \includegraphics[width=.9\columnwidth]{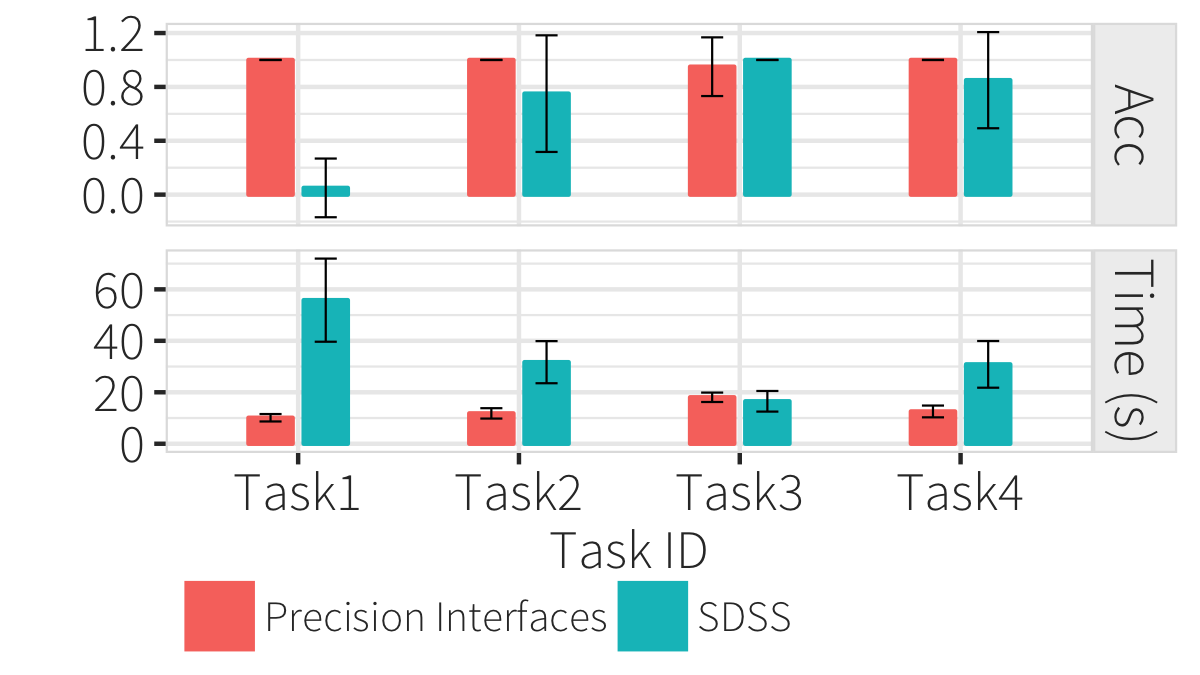}
    \caption{Comparison of response time and accuracy using the original SDSS interface and the automatically generated interface.}
    \label{fig:userstudy}
\end{figure}
Figure ~\ref{fig:userstudy} depicts the average accuracy and time needed for each task for both groups of users. For reference, the average time users needed to perform task 2 (filter objects using the sky coordinates RA (Right Ascension) and dec (Declination) is 34 seconds while it takes only 12 seconds using our generated interface. We explain this by the fact that the original interface does not have default widgets for this task, and users have to choose a combination of options for the widgets to appear and then filter using the widgets, which involves multiple interactions. On the other hand, performing this analysis with our interface requires a single interaction. The case for task 4 (filter using spectrum and redshift) is similar and therefore generated similar results. The response time and accuracy differ the most for task 1 (filtering objects by id) as the original interface does not have a widget for this task and users have to write their own query. Our generated interface led to faster and more accurate analysis for all tasks except task 3 (filter objects by their colors), where both interfaces provided straightforward widgets. Overall, the generated interface created higher quality widgets than the original SDSS interface, which led to an increase in accuracy and a decrease in response time.

\stitle{Interface Preferences}
After users performed the above tasks, we presented them with four interfaces---the SDSS original, \sys, and two manually crafted interfaces---and asked them to choose their preferred interface based on their design.  The aim is to understand the extent to which \sys's interactions are congruent with user expectations.  The two manual interfaces (Figures~\ref{fig:ui-3},~\ref{fig:ui-4}) were implemented by two software engineers that read the SDSS task manual and implemented applications to support the described analyses.

\begin{figure}[h!]
    \centering
    \includegraphics[width=.6\columnwidth]{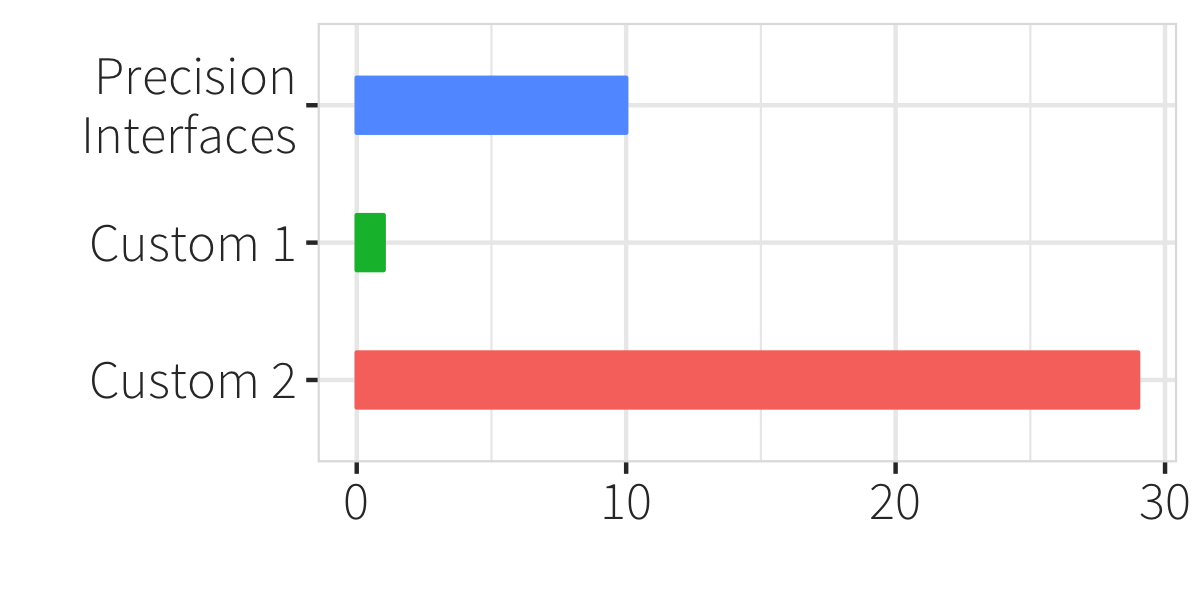}
    \caption{Percentage of user who prefers each interface.}
    \label{fig:preference}
\end{figure}

Figure ~\ref{fig:preference} shows that both \sys and the manually crafted interfaces are preferable to the original. Over 70 percent of the users chose the forth interface while over 20 percent of the users chose \sys.   Out of 40 users, only 1 user chose the first manually designed interface while none chose the original SDSS interface. These results suggest that manually crafted interfaces can vary considerably in perceived quality, and that \sys can generate interfaces competitive with manual implementations.  It also suggests that \sys, without domain-specific knowledge---of the SDSS manuals, the tasks, or the underlaying data---can summarize the salient analysis operations in an interactive interface from query logs that can be simpler to obtain than expert developers.

\subsection{Experiments with Tableau}
\begin{figure*}[t!]
    \centering
    \includegraphics[width=.9\textwidth]{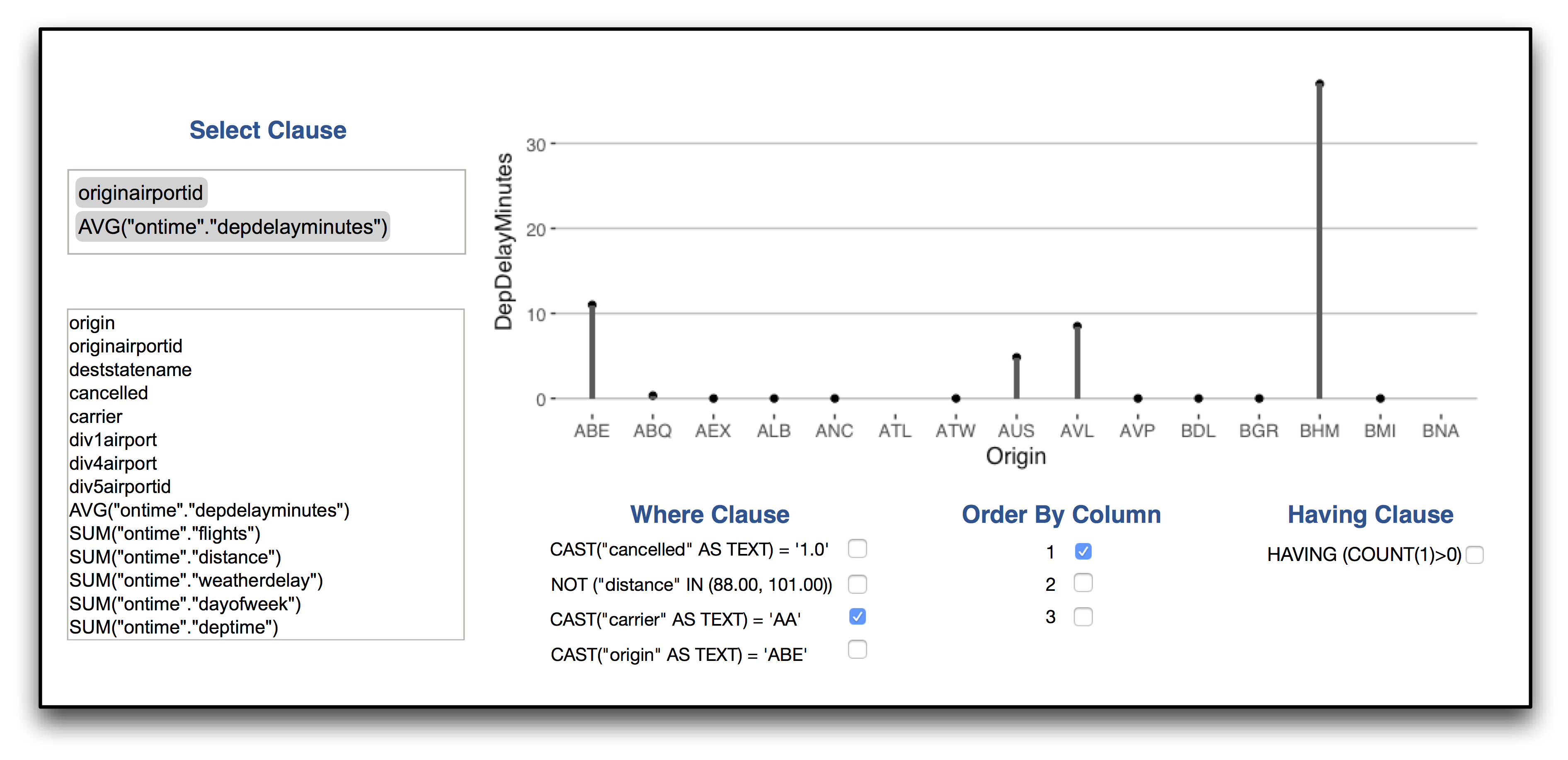}
    \caption{Interface generated for the Tableau log.}
    \label{fig:ui-tableau}
\end{figure*}

In this experiments, we run \sys{} on queries generated directly by Tableau, and check if (1)~our system can detect the underlying interactions and (2)~can generate a simpler, Tableau-like interface. We asked 7 students to use Tableau on the Ontime dataset, using the same setup as that discussed in Section~\ref{sec:case_stud}. We logged the queries generated by Tableau, and obtained $506$ SQL statements ($459$ unique).

 We present the resulting interface in Figure~\ref{fig:ui-tableau}. This UI can express 100\% of the queries in the log, with only 4 components. The select clause widget lets users select the set of attributes and expressions to return (they can drag a column from the bottom box and drop it in the top one). The where clause widget specifies the four predicates that were used; the order by widget shows the three attributes students combined to sort the output. The having clause widget simply adds a no-op expression to the query --- it is a side-effect of how Tableau automatically generates queries, which would likely be removed by the user.

 This use case is ``easy'' for \sys{} because Tableau generates highly structured transformations. In fact, more than 99\% of the edges in the interaction graph express changes of columns in the \texttt{GROUP BY} and the \texttt{SELECT} clause. Those actions correspond to drag and drops in the leftmost component in our interface.
\section{Related Work}


\stitle{User Interface generation:}
Jayapandian et al. automate form-based record search and creation interfaces by analyzing the content of the database~\cite{jayapandian2008automated}. In contrast, we use example queries to synthesize analysis interfaces. In future work, we plan to borrow these ideas and take data and query semantics into account. The UI literature offers a large body of work on model-based interface design~\cite{puerta1994model, vanderdonckt1994automatic, nichols2004improving}, which rely on the developer to provide high level specifications and focus on layout. The above works do not explicitly leverage query logs.

\stitle{Development Libraries:} Tools such as Sikuli~\cite{yeh2009sikuli} or Microsoft Access let non-technical users build their own interfaces. They improve upon lower-level libraries (e.g., Bootstrap) but still require programming and debugging. Similarly, reactive languages (e.g., d3.express~\cite{d3express}, Shiny~\cite{chang2015shiny}, EVE~\cite{eve}, etc) still require programming and are limited to value changes rather than structural program changes.

\stitle{Log Mining:} Historically query log mining has been used in the database literature to detect representative workloads for performance tuning~\cite{chaudhuri1998autoadmin,hellerstein2007architecture}. More recently, it has been used to support data exploration. QueRIE~\cite{eirinaki2014querie} and SnipSuggest~\cite{khoussainova2010snipsuggest} produce context-sensitive suggestions from existing queries at the string level. Query steering~\cite{dimitriadou2014explore} uses a Markov model to produce new statements. Log mining is also extremely common in web search~\cite{silvestri2009mining}, e.g., to augment search results~\cite{hearst2009search}, make suggestions~\cite{cai2016survey} or enable exploration~\cite{chirigati2016knowledge}. \sys exploits and summarizes the structural changes found in query logs into interactive interfaces.

\stitle{Visualization Recommendation:} Visualization recommendation tools such as Panoramic Data~\cite{zgraggen2014panoramicdata}, Zenvisage~\cite{siddiqui2016effortless} and Voyager~\cite{wongsuphasawat2016voyager} constitute a recent and complementary research direction. Those tools help recommend similar data to a given view, while \sys seeks to generate the exploration interface itself.

\stitle{Interface Redesign:}  Interface redesign includes responsive designs that adjust the presentation or selects alternative widgets based on the display size or modality~\cite{adaptive1993}. Similar interface redesign techniques have been used to reduce data entry errors in survey design~\cite{kuangusher}. Those techniques are complementary to ours, which focuses on identifying and selecting task-specific interactions.

\stitle{Programming Languages:} The motivations behind \sys{} are close to those of domain specific languages (DSLs)~\cite{hudak1997domain}, with the major difference that \sys{} targets the interaction domain. A subset of the DSL literature discusses how to build task-specific compilers~\cite{cleaveland1988building}; we will incorporate those ideas in future work.
A related domain of research is \emph{program synthesis}, which seeks to construct programs that satisfy a high level logical description. For instance, Potter's Wheel~\cite{raman2001potter} and Foofah~\cite{jin2017foofah} build data transformation programs based on input and output examples. We target a different problem---\sys{} analyzes query logs, not  input-output pairs.
\section{Conclusion and Discussion}

This paper introduced the use of query logs as the API for interface generation, formalized the problem of extracting and generating interactive interfaces from these logs, and presented \sys as a language-agnostic solution.   To do so, we introduced a unified model over queries, query changes, interfaces, and interactions; and presented algorithms and optimizations to solve the interface generation problem for SQL and SPARQL. Visual interactive interfaces are increasingly relied upon in analysis, and we believe \sys presents an exciting research direction towards improving accessibility for the long-tail of analyses.   From another perspective, we view \sys as a compact {\it visual summary}  of a query log.


We are extending this work in several directions. First is to go beyond syntactic changes and incorporate  metadata, language semantics, database content and other information can lead to richer output interfaces. Second, we will investigate how grammar induction methods~\cite{berwick1987learning} can help us learn or recommend \lang statements directly. Finally, \sys{} currently assumes that \emph{all} queries in the log are relevant to the user's task and in the same language; a third direction is to support complex, multi-language logs.
\\
\stitle{Acknowledgements:} We thank Yifan Wu for the initial inspiration, Anant Bhardwaj for data collection, Laura Rettig on early formulations of the problem, and the support of NSF 1527765 and 1564049.

{\small
\bibliographystyle{abbrv}
\bibliography{main}
\balance
}

\end{document}